\documentclass{amsart}

\RequirePackage{fix-cm}

\usepackage{graphicx}
\usepackage{epstopdf}
\usepackage{latexsym}
\usepackage{amsmath, amssymb,amsfonts,amscd}
\usepackage{mathrsfs}
\usepackage{arydshln}
\usepackage{url}
\usepackage{enumitem}
\usepackage{subfig}
\usepackage{balance}
\usepackage{caption}
\usepackage{color}
\usepackage{natbib}
\usepackage{amsaddr}

\newtheorem{theorem}{Theorem}[section]
\newtheorem{lemma}[theorem]{Lemma}
\newtheorem{proposition}[theorem]{Proposition}
\newtheorem{corollary}[theorem]{Corollary}

\newcommand{\RR}{\mathbb{R}}

\newcommand{\mb}[1][]{\mathbf}
\newcommand{\m}[1]{\mb{m_{#1}}}
\newcommand{\x}{\mb{x}}

\newcommand{\bs}[1][]{\boldsymbol}
\newcommand{\bt}{\boldsymbol{\tau}}
\newcommand{\eqn}{\begin{eqnarray}}
\newcommand{\feqn}{\end{eqnarray}}

\newcommand{\argmin}{\mathop{\rm arg~min}\limits}

\newcommand{\unarydots}{{\scalebox{1}[1.0]{\( \ldots \)}}}
\newcommand{\unaryminus}{{\tiny\scalebox{0.5}[1.0]{\( - \)}}}
\newcommand{\middleminus}{{\scalebox{0.7}[1.0]{\( - \)}}}

\begin{document}

\title{Source localization and denoising: a perspective from the TDOA space
}


\author[M.Compagnoni, A.Canclini, P.Bestagini, F.Antonacci, A.Sarti, S.Tubaro]{Marco Compagnoni}
\address{
Dipartimento di Matematica, Politecnico di Milano \\
            Piazza Leonardo da Vinci 32, 20133, Milan, Italy}
\email{\{marco.compagnoni, antonio.canclini, paolo.bestagini
}
\author[]{Antonio Canclini, Paolo Bestagini, Fabio Antonacci, Augusto Sarti, Stefano Tubaro}
\address{Dipartimento di Elettronica, Informazione e Bioingegneria, Politecnico di Milano \\
            Piazza Leonardo da Vinci 32, 20133, Milan, Italy}
\email{
fabio.antonacci, augusto.sarti, stefano.tubaro \}@polimi.it}

\maketitle

\begin{abstract}
	In this manuscript, we formulate the problem of source localization based on Time Differences of Arrival (TDOAs) in the TDOA space, i.e. the Euclidean space spanned by TDOA measurements.
	More specifically, we show that source localization can be interpreted as a denoising problem of TDOA measurements. As this denoising problem is difficult to solve in general, our analysis shows that it is possible to resort to a relaxed version of it. The solution of the relaxed problem through linear operations in the TDOA space is then discussed, and its analysis leads to a parallelism with other state-of-the-art TDOA denoising algorithms. Additionally, we extend the proposed solution also to the case where only TDOAs between few pairs of microphones within an array have been computed. The reported denoising algorithms are all analytically justified, and numerically tested thorough simulative campaign.
\keywords{TDOA space \and TDOA denoising \and TDOA redundancy \and Source localization}
\end{abstract}

\section{Introduction}\label{sec:intro}
Source localization is a research theme that has significantly grown in popularity in the past few decades, and whose interest ranges from audio to radar. As far as audio signal processing is concerned, several applications including teleconferencing \citep{D_Arca2014}, audio-surveillance \citep{Valenzise2007} and human-machine interaction \citep{Trifa2007} can benefit from the knowledge of the source location. Among the techniques that are available in the literature \citep{Benesty2004}, those based on Time Difference Of Arrival (TDOA) measurements are particularly appreciated for their modest computational requirements. TDOAs, in fact, are usually estimated through peak-picking on the Generalized Cross Correlation of the signals acquired at microphone pairs \citep{Knapp1976,Ianniello1982}, or on the whole set of microphones \citep{Hu2010,Chen2002}. TDOAs can be easily converted to Range Differences (RD), once the sound speed is known. The source location is then found as the point in space that best fits the RD measurements according to properly defined cost functions \citep{Hahn1973,Stoica1988,Schau1987,Huang2001,Beck2008,Schmidt1972}.
More recently, the widespread diffusion of sensor networks stemmed an interest in source localization also in other research communities, such as remote sensing and radar \citep{Koch95,Yimin2008}. In this context range differences are obtained from TDOAs \citep{Kehu2009}, or from energy measurements \citep{Hen2002}. 

The main drawback of TDOA-based localization techniques lies in their sensitivity to measurement noise. In particular, we can distinguish between additive noise (generally due to sampling in the time domain,  circuit noise, but also other physical phenomena) and outlier measurements (produced by reverberation or interfering sources). Outlier identification and removal has been widely studied in the literature (see for instance \citep{Scheuing2008} and references therein; \citep{Canclini2013} and \citep{Canclini2015}). Therefore, applying one of these techniques it is possible to remove outliers from the pool of available measurements. Nonetheless, additive noise still impairs the localization accuracy.

In this manuscript we interpret the problem of source localization studying the effect of additive noise on TDOA measurements using the TDOA space formalism, i.e., a space in which a set of measured TDOAs is mapped into a point. The sensitivity to noise afflicts also Range Differences obtained from energy measurements. Indeed, the hostile propagation conditions yield a difference of the measured energy from the ideal free-field assumption. In the following we will specifically refer to the problem of localizing acoustic sources, but the theory can be readily applied also to other kinds of signals. The concept of TDOA space is not novel and was first introduced in \citep{Spencer2007} for localization purposes. From that representation, a TDOA map (from the space of source locations to the space of TDOAs) was later introduced and analytically derived in \citep{Compagnoni2013a}, which proposed an exhaustive analytic study of the identifiability and invertibility of this map for the three-microphone case. 
In the most general case, given a set of $n+1$ microphones, $q=n(n+1)/2$ TDOAs can be computed considering all the possible microphone pairs. In a noiseless scenario, however, we can always find an independent set of $n$ such TDOAs that we can compute all the other TDOAs from. This is why most TDOA-based algorithms define a reference microphone, with respect to which the $n$ independent TDOAs are computed. In the TDOA space, this corresponds to the fact that TDOAs lie on a linear subspace $V_n$ of the $q$-dimensional TDOA space. This subspace can be computed in closed form through simple considerations. Feasible TDOAs (i.e. points in the TDOA space that correspond to source locations) are bound to lie in a region $\Theta_n \subset V_n$. In \citep{Compagnoni2013a} authors derive $\Theta_n$ in terms of real algebraic geometry. 

Working in the TDOA space essentially means solving an estimation problem in its dual space. As typically done in estimation theory, using a dual domain enables to split a problem into two parts. In our case, using the TDOA space formalism, source localization can be interpreted as a two-step procedure: i) a denoising operation, which consists in removing or attenuating part of the additive noise; ii) the application on the denoised TDOAs of a simple mapping from the TDOA space to the geometric space. Starting from this perspective, in this work we provide a deeper investigation on the geometrical characteristics of the TDOA space.
More specifically, we first derive the correct denoising formulation that fully describes the source localization problem. As the denoising problem formulated this way is not easy to deal with, we resort to a relaxed version of it, which exploits the linear subspace $V_n$. In particular, we show that additive noise can be decomposed into the sum of two orthogonal components, and the relaxed problem formulation aims at reducing only one of them, still positively impacting on source localization.

This relaxed version of the problem was implicitly solved in \citep{Cheung2008} and \citep{Schmidt1996}, where the authors derive closed-form expressions for converting the full set of TDOAs to the nonredundant one. However, authors in \citep{Cheung2008} and \citep{Schmidt1996} limited their analysis to simulations showing that this conversion is able to reduce the impact of noise in localization accuracy. Instead, working in the TDOA space paves the way to a deeper understanding of the impact of relaxed denoising on source localization. In particular we will:
\begin{enumerate}
\item analytically prove the positive effect of denoising on source localization through a set of solid theorems, thus also theoretically validating \citep{Cheung2008} and \citep{Schmidt1996};
\item find a solution to the relaxed denoising problem when some TDOA measurements are not available;
\item quantify analytically the improvement in localization accuracy brought by the use of the denoised TDOAs given a specific localization algorithm in use. We accomplish this analysis by means of the error propagation theory introduced in \citep{Compagnoni2012}.
\end{enumerate}

We test the presented algorithm also under different noise hypotheses to show that it works also if the underlying assumptions are not strictly verified. In particular, Monte Carlo simulations were carried out to show how different state-of-the-art techniques (the SRD-LS algorithm \citep{Beck2008}, Least Squares \citep{Smith1987}) and Gillette-Silverman \citep{Gillette2008a}) methods benefit from denoising, approaching the RMSE Lower Bound (RLB) implied by the Cramer-Rao Lower Bound (CRLB) \citep{Benesty2004}. We also show that it is possible to perform denoising on a set of TDOAs including $q-s, ~s>0$ measurements, with an apparent advantage in terms of accuracy.

The rest of the manuscript is structured as follows. In Section~\ref{sec:TS} we deeply introduce the formalism of TDOA space. In Section~\ref{sec:source_loc} we interpret the problem of source localization within this context. In Section \ref{sec:denoising} we provide the denoising formulation of the source localization problem, also reporting the relaxed problem version and an algorithm for its solution. A parallelism with related state-of-the-art works is also provided. In Section~\ref{impact_on_loc} we analytically prove the positive impact of denoising on source localization, and provide additional simulative analysis. Section~\ref{sec:denoisinglTDOA} is devoted to denoising problem formulation and solution in case some TDOAs are missing within the pool of measurements (i.e., we measure TDOAs using only a few pairs of microphones). Finally, Section~\ref{sec:conclusions} remarks some final conclusions highlighting possible open future research lines.

\section{Theoretical background}\label{sec:TS}
In this section we offer the reader some background that will simplify the reading of this article. In particular, we first provide the formal definition of the TDOA space. Then, we give the interpretation of noisy measurements and source localization problem in the TDOA space.

\subsection{The TDOA space}
The ideas of the TDOA space; the feasible set of TDOA measurements; and the TDOA map appeared in several manuscripts concerning multilateration, see for example \citep{Schmidt1996,Grafarend2002,Spencer2007,Compagnoni2013a}.
These concepts are the essential ingredients for the mathematical definition and analysis of many problems involving TDOA measurements, such as source localization, synchronization and calibration of the receivers.
A recent example in this direction can be found in \citep{Pineda2014}, where the TDOA space formalism is used for defining a novel algorithm to estimate the  TDOAs and concurrently locate the source.
In the following, we present the basic definitions and properties regarding the TDOA space.

Let $\m{i}=(x_i,y_i,z_i)^T,\ i=0,\ldots,n$ be the sensor locations and $\x$ be the source position in the 3D Euclidean space $\RR^3$. For notational simplicity, and with no loss of generality, in what follows we assume the sound speed to be equal to 1, so that the noiseless TDOAs correspond to the range differences. This way, given any pair of sensors $(\m{j},\m{i})$, $n\geq j>i\geq 0$, the relative TDOA is a function of the source position $\x$ and it can be defined as
\begin{equation}\label{eq:TDOAs1}
\begin{array}{cccc}
\tau_{ji}: & \RR^3 & \longrightarrow & \RR\\
& \x & \longmapsto & \tau_{ji}(\x)
\end{array},
\end{equation}
where
\begin{equation}\label{eq:TDOAs2}
\tau_{ji}(\x)=\Vert\x-\m{j}\Vert-\Vert\x-\m{i}\Vert.
\end{equation}
If we collect the $q=\frac{n(n+1)}{2}$ range differences in a $q$--dimensional vector, we obtain the map
\begin{equation}\label{eq:TDOAmap}\
\begin{array}{cccc}
\boldsymbol{\tau_n^*}: & \RR^3 & \longrightarrow & \RR^q\\
& \x & \longmapsto & (\tau_{10}(\x),\tau_{20}(\x),\ldots,\tau_{n\,n-1}(\x))^T
\end{array}\ .
\end{equation}
In \citep{Compagnoni2013a}, $\boldsymbol{\tau_n^*}$ has been called the complete TDOA map, while the resulting target set $\RR^q$ of $\boldsymbol{\tau_n^*}$ is referred to as the TDOA space or $\tau$--space. 
Clearly, a point in the TDOA space corresponds to any set of TDOA measurements. Moreover, in a noiseless scenario, the subset of the $\tau$--space containing the TDOAs generated by all the potential source positions coincides with the image $\text{Im}(\boldsymbol{\tau_n^*})$ of the TDOA map, and we call it $\Theta_n$. This means that any collection of noiseless TDOAs defines a point $\boldsymbol{{\tau}}= (\tau_{10},\ldots,\tau_{n\,n-1})^T\in\Theta_n$ and viceversa.

\subsection{The 2D case with n=2}\label{sec:D2n2}
The study of the properties of $\boldsymbol{\tau_n^*}$ is a fundamental step towards a deeper understanding of the geometrical acoustics model for TDOA--based localization. However, since its inherent complexity, the full description of the general case of $\boldsymbol{\tau_n^*}$ goes beyond the scope of this manuscript. In this section, we summarize the main results contained in \citep{Compagnoni2013a,Compagnoni2013b} on the minimal case of two dimensional source localization, with three synchronized and calibrated sensors.\footnote{In order to simplify the presentation, we consider only the case with the microphones in general position on the plane, i.e. they do not lie on a line. The interested reader can find the complete analysis for every scenario and the proofs in the original manuscripts.}
We report this analysis because this is the minimal non trivial case of TDOA-based localization, the only one that has been exhaustively studied and where one may observe some important features characterizing every localization model. We will return on this model in Section \ref{sec:denoising}.

In the planar case, the set $\Theta_2$ is a surface embedded into $\RR^3$, being the image of the restriction of the TDOA map $\boldsymbol{\tau_2^*}$ to $\RR^2$ (by abuse of notation, we continue to name it $\boldsymbol{\tau_2^*}$). Actually, one can interpret $\boldsymbol{\tau_2^*}$ as a (radical) parameterization of $\Theta_2.$ Moreover, it is well known that the three TDOAs are not independent, since they satisfy the zero-sum condition (ZSC) \citep{Scheuing2006}. Indeed, the linear relation $ \tau_{21}(\x ) = \tau_{20}(\x ) - \tau_{10}(\x ) $ holds for each $ \x \in \RR^2.$ Geometrically speaking, this means that three noiseless TDOAs are constrained on the plane
\begin{equation}
V_2=\{\bs{\tau}\in\RR^3\ |\ \tau_{10}-\tau_{20}+\tau_{21}=0\}\subset\RR^3
\end{equation}
and so $\Theta_2\subseteq V_2$.

\begin{figure}[htb]
\begin{center}
\resizebox{6.5cm}{!}{
  \includegraphics{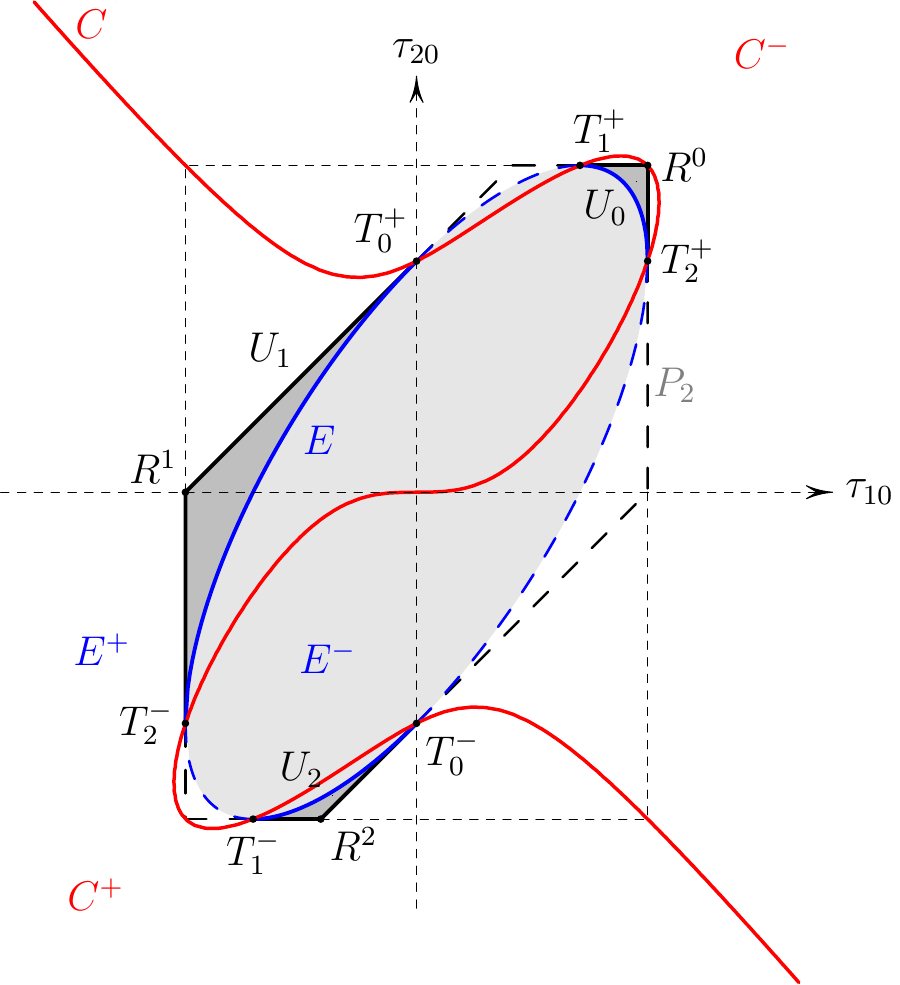}}
  \caption{\label{fig:tauimage}
The image of $\bs{\tau_2}$ is the gray subset of the hexagon $P_2$ with continuous and dashed sides. In the light gray region $E^-$ the map $\bs{\tau_2}$ is $ 1$--to--$1,$ while in the medium gray region $U_0 \cup U_1 \cup U_2$ the map $\bs{\tau_2}$ is $2$--to--$1.$ Let us observe that $U_0 \cup U_1 \cup U_2\subset C^+\cap P_2.$ The continuous part of the boundary of the hexagon and the blue ellipse $E,$ together with the vertices $R^i,$ are in the image, and there $\bs{\tau_2}$ is $ 1$--to--$1.$ The points $T_i^\pm$ and the dashed boundaries do not belong to Im($\bs{\tau_2}$).}
\end{center}
\end{figure}

\begin{figure}[htb]
\begin{center}
\resizebox{7cm}{!}{
\includegraphics{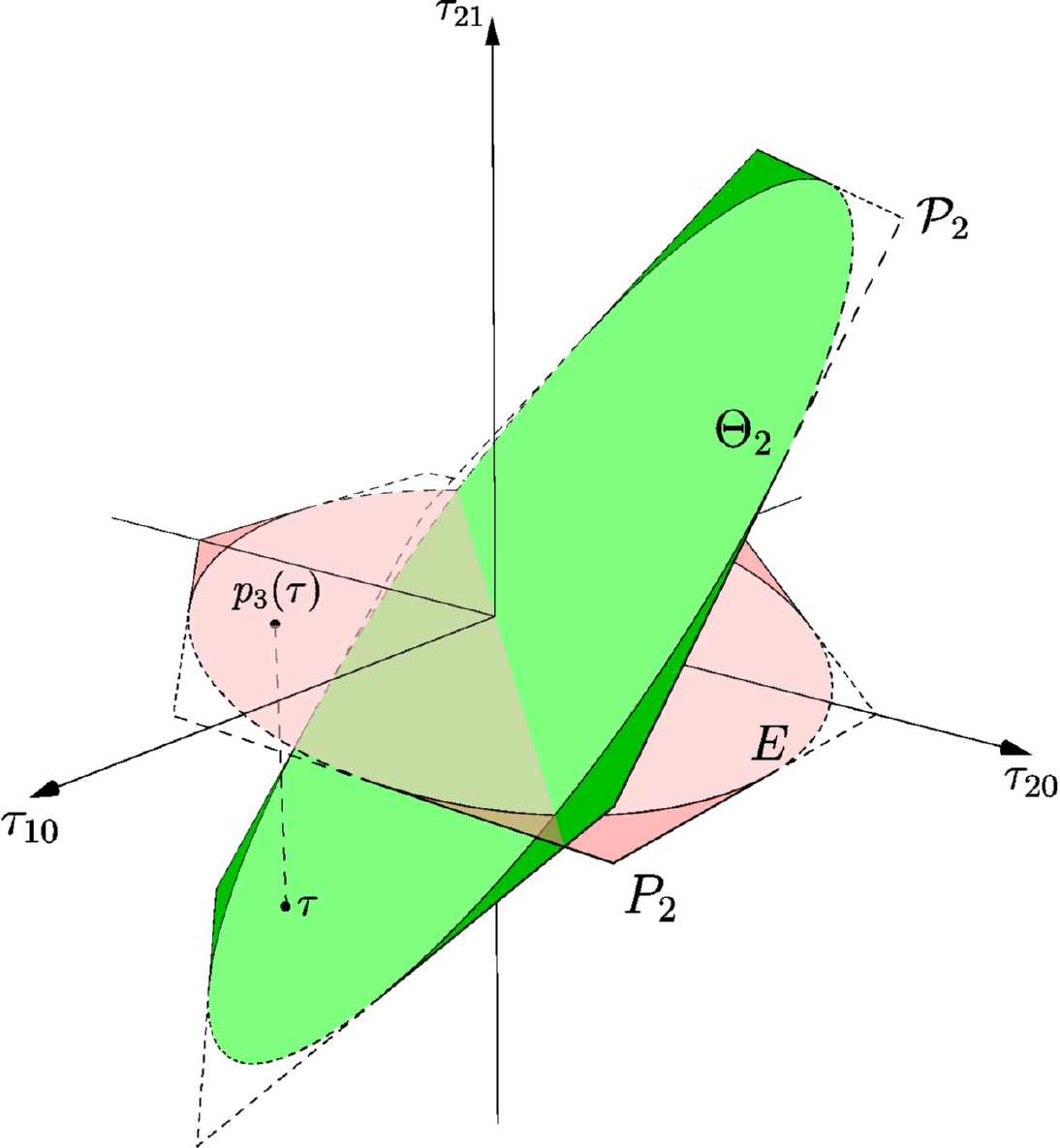}}
\caption{\label{fig:taucomimage}The image of $\bs{\tau_2^*}$ is the green subset of the hexagon $\mathcal{P}_2\subset V_2,$ while the image of $ \bs{\tau_2}$ is the red subset of $P_2.$ There is a 1--to--1 correspondence between Im($\bs{\tau_2^*}$) and Im($\bs{\tau_2}$) via the projection map $p_3$.   In the lightly shaded regions, the TDOA maps are $1$--to--$1$, while in the more darkly shaded regions the maps are $2$--to--$1$.}
\end{center}
\end{figure}

Because of the above linear relation, in the literature it is customary to work with a reference microphone, for example $\m{0},$ and to consider only the two TDOAs $\tau_{10}(\x),\tau_{20}(\x).$ Mathematically speaking, let us define the reduced TDOA map
\begin{equation}
\begin{array}{cccc}
\bs{\tau_2}: & \RR^2          & \longrightarrow & \RR^2\\
 & \x   & \longrightarrow & \quad
(\tau_{10}(\x),\tau_{20}(\x))
\end{array}
\end{equation}
and let us consider the projection map $p_3:\RR^3\rightarrow\RR^2$ forgetting the third coordinate $\tau_{21}$ of the $\tau$--space. Then, we have $\bs{\tau_2}=p_3\circ\bs{\tau_2^*}$ and $p_3$ is a natural bijection between $\text{Im}(\bs{\tau_2^*})$ and $\text{Im}(\bs{\tau_2}).$ Hence, one can investigate the properties of the noiseless TDOA model by studying the simpler map $\bs{\tau_2}$. For the sake of simplicity, in Figure \ref{fig:tauimage} we draw $\text{Im}(\bs{\tau_2})$ for the configuration of the microphone at $\m{0}=(0,0)^T,\ \m{1}=(1,0)^T\ \text{and}\ \m{2}=(1,1)^T$. Symbols defined therein are introduced in the next few paragraphs. Figure \ref{fig:taucomimage} shows its relation with $\text{Im}(\bs{\tau_2^*}).$

Let us define the displacement vectors $\mb{d_{ji}}=\m{j}-\m{i}$; their Euclidean norms $d_{ji}=\Vert\mb{d_{ji}}\Vert,\ i,j=0,1,2$; the scalar
$
W=\det\left(
\begin{array}{ccc}
\mb{d_{10}} &\ & \mb{d_{20}}
\end{array}\right)
$; and the matrix
$$
\mb{H}=\left(
\begin{array}{cc}
0 & -1\\
1 & 0
\end{array}\right).
$$
First of all, Im$(\bs{\tau_2})$ is contained into the hexagon $P_2$ defined by the triangle inequalities:
\begin{equation}\label{eq:politopo}
\left\{ \begin{array}{l}
-d_{10} \leq \tau_{10} \leq d_{10} \\
-d_{20} \leq \tau_{20} \leq d_{20} \\
-d_{21} \leq \tau_{20} - \tau_{10} \leq d_{21}
\end{array} \right. .
\end{equation}
In particular, the vertices $R^0=(d_{10},d_{20}),\ R^1=(-d_{10},d_{21}-d_{10}),\ R^2=(d_{21}-d_{20},-d_{20})$ of $P_2$ correspond to the pairs of TDOAs associated to a source at $\m{0},\m{1},\m{2},$ respectively. Then, by following the analysis contained in Section 6 of \citep{Compagnoni2013a}, for any $\bt=(\tau_{10},\tau_{20})\in\RR^2$ we define the vectors
\begin{equation}
\begin{array}{l}
\mb{v}(\bt)=\mb{H}\,(\tau_{20}\, \mb{d_{10}} - \tau_{10}\, \mb{d_{20}})\,,\qquad 
\mb{l_0}(\bt)=\displaystyle
\mb{H}\,\frac{(d_{20}^2-\tau_{20}^2)\,\mb{d_{10}} -
(d_{10}^2-\tau_{10}^2)\,\mb{d_{20}}}
{2\, W}
\end{array}
\end{equation}
and the polynomials
\begin{equation}
a(\bt) = \Vert\mb{v}(\bt)\Vert^2 -W^2,\qquad
b(\bt) = \mb{v}(\bt)^T\cdot\mb{l_0}(\bt),\qquad
c(\bt) = \Vert\mb{l_0}(\bt)\Vert^2.
\end{equation}
$\text{Im}(\bs{\tau_2})$ and the admissible source positions $\bs{\tau_2}^{-1}(\bt)$ can be computed in terms of these polynomials.

We have the following facts:
\begin{itemize}
\item
$a(\bt)=0$ defines the unique ellipse $E$ tangent to every facet of $P_2.$ We name $E^-$ and $E^+$ the interior and the exterior regions of $E$ where $a(\bt)<0$ and $a(\bt)>0,$ respectively;
\item
$b(\bt)=0$ defines a cubic curve $C.$ We name $C^-$ and $C^+$ the regions where $b(\bt)<0$ and $b(\bt)>0,$ respectively;
\item
$c(\bt)$ is a quartic non negative polynomial.
\end{itemize}
In \citep{Compagnoni2013a} it has been proved that the image of $\bt_2$ is the set
\begin{equation}
\mbox{Im}(\bs{\tau_2}) = E^- \cup (C^+\cap P_2) \cup R^0.
\end{equation}
For each $\bt\in\mbox{Im}(\bs{\tau_2})$ we have at most two admissible source positions, whose coordinates are given by the formula
\begin{equation}\label{eq:inv-image}
\x_\pm(\bt) = \m{0}+\mb{l_0}(\bt) + \lambda_\pm(\bt) \mb{v}(\bt),
\end{equation}
where $ \lambda_\pm(\bt) $ are the solutions of the quadratic equation $a(\bt)\lambda^2+2b(\bt)\lambda+c(\bt)=0:$
$$
\lambda_{\pm}(\bt)=\frac{-b(\bt)\pm\sqrt{b(\bt)^2-a(\bt)c(\bt)}}{a(\bt)}\,.
$$
For $\bt\in (E^-\cup\,E\,\cup\partial P_2)\cap\mbox{Im}(\bs{\tau_2})$ we have to take only the $\x_+(\bt)$ solution and so we have uniqueness of localization. On the complementary set, that is the union of the three disjoint sets $U_0,\ U_1$ and $U_2$ depicted in medium gray in Figure \ref{fig:tauimage}, the map $\bs{\tau_2}$ is $2$--to--$1$ and there is an intrinsic ambiguity in the source position between the two solutions $\x_\pm(\bt)$.

\begin{figure}[htb]
\begin{center}
\resizebox{5cm}{!}{
  \includegraphics{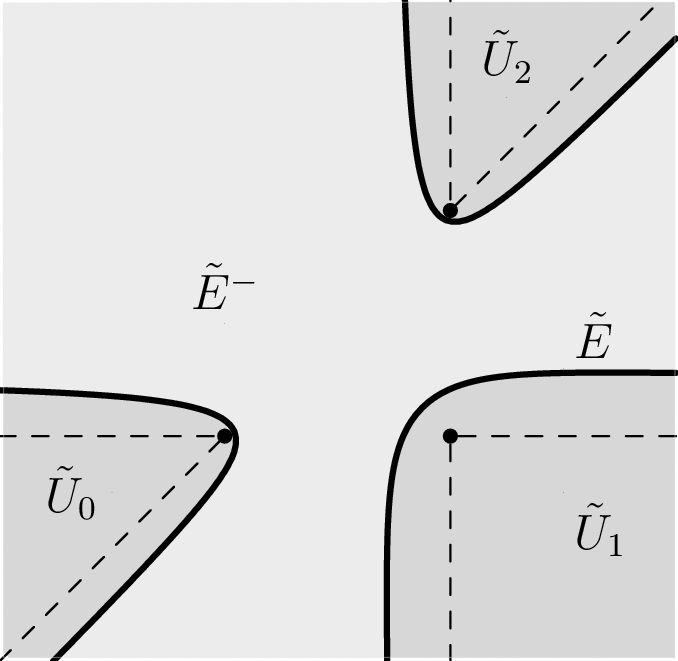}}
\caption{\label{fig:x-plane}The different localization regions and the curve $ \tilde{E} $ in the $x$--plane. The microphones are the marked points $ \m{0}=(0,0),\ \m{1}=(1,0)$ and $\m{2}=(1,1).$ The curve $\tilde{E}$ separates the light gray region $\tilde{E}^-$, where the map $\bt_2$ is 1--1 and it is possible to locate the source, and the medium gray region $\tilde{U}_0\cup\tilde{U}_1\cup\tilde{U}_2$, where $\bs{\tau_2}$ is 2--1 and the localization is not unique. On the dashed lines the localization is possible but very sensitive to the measurement noise.}
\end{center}
\end{figure}

For the sake of completeness, in Figure \ref{fig:x-plane} we depict the corresponding localization regions in the $x$--plane. Roughly speaking, we have the preimage of the interior of the ellipse $ \tilde{E}^- =\bs{\tau_2}^{-1}(E^-)$, where the TDOA map is $1$--to--$1$ and the source localization is possible, and the preimages $ \tilde{U}_i=\bs{\tau_2}^{-1}(U_i)$, for $ i=0,1,2,$ where the map is $2$--to--$1$ and there is no way to uniquely locate the source. The region of transition is the \emph{bifurcation curve} $\tilde{E} =\bs{\tau_2}^{-1}(E),$ that is a quintic algebraic curve \citep{Compagnoni2013b} consisting of three disjoint and unbounded arcs, one for each arc of $ E $ contained in $ \mbox{Im}(\bs{\tau_2}).$ As a point $\bt$ in one of the $U_i$ gets close to $E$, the solution $\x_+(\bt)$ gets close to a point on $\tilde{E}$, while $\x_-(\bt)$ goes to infinity. The sets $\tilde{E}^-, \tilde{U}_0, \tilde{U}_1, \tilde{U}_2 $ are open subsets of the $ x$--plane, separated by the three arcs of $ \tilde{E} $.

Finally, the union $D$ of the six dashed half--lines outgoing from the receivers is the \emph{degeneracy locus} of the TDOA map, where the rank of the Jacobian matrix of $\bs{\tau_2}$ drops. $D$ is the preimage of the six segments in $\partial P_2\cap\text{Im}(\bs{\tau_2}).$ On $D$ the two solutions $\x_\pm(\bt)$ are coincident, thus the TDOA map is $1$--to--$1.$ Furthermore, $D$ divide each $ \tilde{U}_i $ into two connected components and $\bt_2$ is a bijection between each of them and the corresponding $U_i$.

\subsection{The general case}\label{sec:generalTDOAspace}
As we said above, some of the properties we described in the minimal planar case are common to every TDOA-based localization model. In particular, we have the following proposition.
\begin{proposition}\label{prop:Vn}
	Let us take $n+1$ sensors at $\m{0},\ldots,\m{n}$ in $\RR^3$, where $n\geq 2.$ Then,
	$\Theta_n$ is a subset of the $n$--dimensional linear subspace $V_n\subset\RR^q$ defined by equations
	\begin{equation}\label{eq:lineq}
	-\tau_{i0}+\tau_{j0}-\tau_{ji}=0,\qquad 0<i<j\leq n\;,
	\end{equation}
	representing the ZSCs for all the microphone triplets containing $\m{0}$.
\end{proposition}
\noindent\emph{Proof:}
in a configuration with $n+1$ microphones, the maximum number of independent TDOAs is equal to $n.$ In particular, if we take $\m{0}$ as the reference microphone, the $n$ TDOAs $\{\tau_{10}(\x),\ldots,\tau_{n0}(\x)\}$ are independent, while the others satisfy equations \eqref{eq:lineq}, as can be easily verified using definition \eqref{eq:TDOAs2}. These are $q-n$ independent homogeneous linear equations, therefore they define an $n$--dimensional linear subspace $V_n$ of the $\tau$--space and $\Theta_n$ is a subset of $V_n$.
\hfill$\square$\vspace{1mm}

This means that, also in the general case, the set of feasible TDOAs is contained into an $n$--dimensional linear subspace of the TDOA space $\RR^q$. This property stays at the basis of the denoising procedure that we describe in the next sections. However, $\Theta_n$ is strictly contained in $V_n.$ Indeed, since $\Theta_n$ is the image of $\RR^3$ through the almost everywhere smooth function $\bs{\tau_n},$ its dimension is equal to 3. As above, one can consider $\bs{\tau_n}$ as a radical parameterization of $\Theta_n.$ This way the feasible set becomes a topological manifold (possibly with a boundary, as for $\Theta_2$), that is almost everywhere differentiable. Moreover, we can reasonably conjecture that $\Theta_n$ can be described again in terms of algebraic equations and inequalities, hence it is a so called semialgebraic variety \citep{Basu2006}.
\section{Interpretation of source localization in the TDOA space}\label{sec:source_loc}
The TDOA space formalism that we just introduced can be used to provide a geometric interpretation of TDOA-based source localization problem. As a matter of fact, other geometric interpretations have been proposed in the literature \citep{Bestagini2013}. However, working in the TDOA space also highlights that source localization can be solved as a TDOA denoising problem. In this section, we provide such interpretation of source localization in the TDOA space, starting from a commonly used statistical noise model.

\subsection{Statistical noise model}
In the presence of measurement errors, we must resort to statistical modeling. In this manuscript we assume the TDOAs associated to a source in $\x$ to be described by
\begin{equation}\label{eq:TDOAstatmod}
\bs{\hat{\tau}_n^*}(\x)=\bs{\tau_n^*}(\x)+\bs{\varepsilon},
\end{equation}
where $\bs{\varepsilon}\sim N(\mb{0},\bs\Sigma)$ is an additive Gaussian noise. Also techniques that compute Range Differences from other measurements, such as energy, are prone to additive noise. In this latter case, in particular, the magnitude of the additive noise becomes relevant. Under the assumption in \eqref{eq:TDOAstatmod}, the probability density function (p.d.f.) of the TDOA set is \citep{Benesty2004}
\begin{equation}\label{eq:TDOAprob}
p(\bs{\hat{\tau}};\bs{\tau_n^*}(\mb{x}),\bs\Sigma)=
\frac{e^{-\frac{1}{2}(\bs{\hat{\tau}}-\bs{\tau_n^*}(\mb{x}))^T \bs\Sigma^{-1}(\bs{\hat{\tau}}-\bs{\tau_n^*}(\mb{x}))}}{\sqrt{(2\pi)^q|\bs\Sigma|}}\,.
\end{equation}
Since the covariance matrix $\bs\Sigma$ is symmetric and positive defined, from a geometric standpoint (see, for example, \citep{Amari2000}) the Fisher matrix $\bs\Sigma^{-1}$ defines a scalar product on $\RR^q$:
\begin{equation}\label{eq:Fisherscalarprod}
\langle\mb{v_1},\mb{v_2}\rangle_{\bs\Sigma^{-1}} = \mb{v_1}^T \bs\Sigma^{-1} \mb{v_2},\qquad \mb{v_1},\mb{v_2}\in\RR^q \; .
\end{equation}
This way, the TDOA space turns out to be equipped with a Euclidean structure, whose distance is known in the statistical literature as the Mahalanobis distance:
\begin{equation}\label{eq:Mahalanobis}
\Vert\mb{v}\Vert_{\bs\Sigma^{\unaryminus 1}} =\sqrt{\mb{v}^T \bs\Sigma^{-1} \mb{v}}\,,\qquad \mb{v}\in\RR^q \; .
\end{equation}
With this setting, the p.d.f. \eqref{eq:TDOAprob} can be rewritten as
\begin{equation}\label{eq:TDOAprob2}
p(\bs{\hat{\tau}};\bs{\tau_n^*}(\mb{x}),\bs\Sigma)=
\frac{e^{-\frac{1}{2}\Vert\bs{\hat{\tau}}-\bs{\tau_n^*}(\mb{x})
		\Vert_{\bs\Sigma^{\unaryminus 1}}^2}}{\sqrt{(2\pi)^q|\bs\Sigma|}}\,,
\end{equation}
which depends only on the Mahalanobis distance between $\bs{\hat{\tau}}$ and $\bs{\tau_n^*}(\mb{x}).$

\subsection{Source localization}
The first application of the TDOA space and map was in the study of the TDOA--based source localization (see \citep{Spencer2007,Compagnoni2013a}). As a matter of fact, the fundamental questions in localization problems can be readily formulated in terms of $\bs{\tau_n^*}.$ In a noiseless scenario, the analysis of the existence and uniqueness of localization is equivalent to the study of the set $\Theta_n$ and the invertibility of $\bs{\tau_n^*}$. As we saw in Section \ref{sec:D2n2} for the minimal planar case, for a given $\bs\tau\in\RR^q$ there exists a unique source at position $\bs{\tau_n^*}^{-1}(\bs\tau)$ if, and only if, $\bs\tau$ is a point lying on a region of $\Theta_n$ where the TDOA map is $1$--to--$1.$

In case of noisy measurements, the data errors force the point $\bs{\hat{\tau}}$ not to lie on $\Theta_n.$ Therefore, to localize the source one needs an estimation procedure. Most of the algorithms proposed in the literature rely on the optimization of a cost function, based on some criterion that can be either statistically motivated (e.g., Maximum-Likelihood estimation \citep{Benesty2004}) or not (e.g., linear Least Squares \citep{Smith1987,Gillette2008a}, Squared Range-Differences-based least squares estimation \citep{Beck2008}, etc.). The source position is thus found as the point $\mb{\bar x}$ that minimizes a suitable non-negative cost function $f(\bs{\hat\tau},\x)$, defined so that its value is zero in noiseless conditions, i.e.,
\begin{equation}
\label{eq:noiseless_loc}
f(\bs{\tau_n^*}(\x),\x)=0
\end{equation}
for every $\x\in\RR^3.$ In mathematical terms, source localization is therefore formulated as
\begin{equation}\label{eq:generic_source_loc}
\mb{\bar x} =
\underset{\x\in\RR^3}{\text{argmin}}\
f(\bs{\hat\tau},\x)\;.
\end{equation}

In the TDOA space, the estimated source position is associated to the feasible point $\bs{\bar\tau}=\bs{\tau_n^*}(\mb{\bar x})\in\Theta_n$. Therefore, any localization algorithm maps a noisy TDOA vector $\bs{\hat\tau}$ onto a feasible TDOA vector $\bs{\bar\tau}$. It is worth noticing that different algorithms may produce different source estimates, corresponding to mappings onto likewise different feasible TDOA vectors. A special case occurs when the input TDOA vector is feasible, i.e., when $\bs{\hat\tau}\in\Theta_n$. In this case, there exists a point $\mb{\bar x}$ so that $\bs{\hat\tau}=\bs{\tau_n^*}(\mb{\bar x})$. Thus, in virtue of \eqref{eq:noiseless_loc}, we expect any algorithm to produce the same estimate $\mb{\bar x} = \bs{\tau_n^*}^{-1}(\bs{\hat\tau})$. 

Let us now focus on the Maximum-Likelihood (ML) estimator, which is known to be optimal in the statistical sense. For the Gaussian noise model described above, the ML localization problem can be formulated as \citep{Benesty2004}
\begin{equation}\label{eq:ML_1}
\mb{\bar x}_{\text{\tiny ML}} = \underset{\x\in\RR^3}{\text{argmax}}\
\displaystyle
\frac{e^{-\frac{1}{2}\Vert\bs{\hat{\tau}}-\bs{\tau_n^*}(\mb{x})
		\Vert_{\bs\Sigma^{\unaryminus 1}}^2}}{\sqrt{(2\pi)^q|\bs\Sigma|}}\;.
\end{equation}
By defining $f_{\text{\tiny ML}}(\bs{\hat\tau},\x) = 
\Vert\bs{\hat{\tau}}-\bs{\tau_n^*}(\mb{x})\Vert_{\bs\Sigma^{\unaryminus 1}}^2,$ we have
\begin{equation}\label{eq:ML_2}
\bar{\x}_{\text{\tiny ML}} = \underset{\x\in\RR^3}{\text{argmin}}\ f_{\text{\tiny ML}}(\bs{\hat\tau},\x).
\end{equation}
In the TDOA space framework, the ML estimator has a neat geometric interpretation. Indeed, solving the ML problem is equivalent to finding the point $\bs{\bar\tau}_{\text{\tiny ML}}=\bs{\tau_n^*}(\mb{\bar x}_{\text{\tiny ML}})\in\Theta_n$ at minimum Mahalanobis distance from $\bs{\hat{\tau}}$:
\begin{equation}\label{eq:MLE-TDOAspace2}
\bs{\bar{\tau}}_{\text{\tiny ML}} =
\underset{\bt\in\Theta_n}{\text{argmin}}\
\Vert\bs{\hat{\tau}}-\bt\Vert_{\bs\Sigma^{\unaryminus 1}}^2.
\end{equation}
It trivially follows that the source position $\mb{\bar x}$ estimated by means of a generic localization algorithm leads to a point $\bs{\bar\tau} = \bs{\tau_n^*}(\mb{\bar x})\in\Theta_n$ such that
\begin{equation}
\Vert\bs{\hat{\tau}}-\bs{\bar\tau}\Vert_{\bs\Sigma^{\unaryminus 1}} \geq \Vert\bs{\hat{\tau}}-\bs{\bar\tau}_{\text{\tiny ML}}\Vert_{\bs\Sigma^{\unaryminus 1}}\;.
\end{equation}
In other words, the distance $\Vert\bs{\hat{\tau}}-\bs{\bar\tau}\Vert_{\bs\Sigma^{\unaryminus 1}}$ is bounded from below by the one obtained through ML estimation.

In the light of the above considerations, we can interpret source localization in the TDOA space as a two-step procedure:
\begin{enumerate}
	\item mapping of the noisy TDOAs $\bs{\hat{\tau}}$ onto the corresponding feasible vector $\bs{\bar\tau}\in\Theta_n$, accordingly with the chosen optimization criterion;
	\item recovering of the estimated source position as $\mb{\bar x} = \bs{\tau_n^*}^{-1}(\bs{\bar\tau})$.
\end{enumerate}
Note that step 1 represents a TDOA denoising operation. Moreover, once this step has been accomplished, step 2 is straightforward when the mapping $\bs{\tau_n^*}$ is 1-to-1, that is the standard case for a sufficiently large number $n$ of sensors in general positions. From this perspective, source localization can be considered as a TDOA denoising problem.
\section{Denoising of TDOAs}\label{sec:denoising}
As discussed in Section \ref{sec:source_loc}, source localization can be solved in the TDOA space as a denoising problem. Given a noisy TDOA vector $\bs{\hat{\tau}}$, this corresponds to finding a feasible (i.e., denoised) TDOA vector belonging to $\Theta_n$ according to some criteria that depends on the chosen cost function $f(\bs{\hat\tau},\x)$. If we consider the Maximum-Likelihood formulation of the problem, the criterion is readily formulated as in \eqref{eq:MLE-TDOAspace2}. Therefore, statistically speaking, the best achievable denoising corresponds to finding the feasible TDOA vector at the minimum Mahalanobis distance from the noisy one.

Despite the error function $\Vert\bs{\hat{\tau}}-\bt\Vert_{\bs\Sigma^{\unaryminus 1}}^2$ could suggest a standard weighted linear least-squares solution \citep{Teunissen2000}, the problem in \eqref{eq:MLE-TDOAspace2} can not be solved in a linear fashion. Indeed, the search space is $\Theta_n$, which turns \eqref{eq:MLE-TDOAspace2} in a difficult, in general non convex, optimization problem.
However, in the following we investigate the possibility of relaxing the problem in \eqref{eq:MLE-TDOAspace2}, leveraging on the fact that $\Theta_n$ is contained in the linear subspace $V_n$ of the TDOA space.

\subsection{From the complete to the relaxed denoising problem}
\label{sec:relaxed_den_prob}
We claimed above that solving the ideal denoising problem \eqref{eq:MLE-TDOAspace2} is a non trivial task. However, by working in the TDOA space we can subdivide such problem in two distinct steps. Let us consider the orthogonal projection $\mathcal{P}(\bs{\hat{\tau}};\bs\Sigma)$ of $\bs{\hat{\tau}}\in\RR^q$ on $V_n,$ with respect to the scalar product $\langle\ ,\ \rangle_{\bs\Sigma^{-1}}$. From the general properties of this map, we have the following equivalences:
$$
\begin{array}{lcl}
\bs{\bar{\tau}}_{\text{\tiny ML}} & = &
\underset{\bt\in\Theta_n}{\text{argmin}}\
\Vert\bs{\hat{\tau}}-\bt\Vert_{\bs\Sigma^{\unaryminus 1}}^2\\
& = & \underset{\bt\in\Theta_n}{\text{argmin}}
\Vert(\bs{\hat{\tau}} \middleminus \mathcal{P}(\bs{\hat{\tau}};\bs\Sigma))+
({P}(\bs{\hat{\tau}};\bs\Sigma) \middleminus \bt)
\Vert_{\bs\Sigma^{\unaryminus 1}}^2\\
& = & \underset{\bt\in\Theta_n}{\text{argmin}}\left(
\Vert\bs{\hat{\tau}} \middleminus \mathcal{P}(\bs{\hat{\tau}};\bs\Sigma)
\Vert_{\bs\Sigma^{\unaryminus 1}}^2+
\Vert\mathcal{P}(\bs{\hat{\tau}};\bs\Sigma) \middleminus \bt\Vert_{\bs\Sigma^{\unaryminus 1}}^2\right)\\
& = & \underset{\bt\in\Theta_n}{\text{argmin}}\
\Vert\mathcal{P}(\bs{\hat{\tau}};\bs\Sigma) \middleminus \bt\Vert_{\bs\Sigma^{\unaryminus 1}}^2\,,
\end{array}
$$
where in the third equality we used the fact that $\bs{\hat{\tau}}-\mathcal{P}(\bs{\hat{\tau}};\bs\Sigma)$
and $\mathcal{P}(\bs{\hat{\tau}};\bs\Sigma)-\bt$ are orthogonal each other, thus
$$
\langle\bs{\hat{\tau}}-\mathcal{P}(\bs{\hat{\tau}};\bs\Sigma),
\mathcal{P}(\bs{\hat{\tau}};\bs\Sigma)-\bs{\bar{\tau}}_{\text{\tiny ML}}\rangle_{\bs{\Sigma}^{\unaryminus 1}}=0\,.
$$
This means that the ML estimation gives exactly the same results if we start from the original data $\bs{\hat{\tau}}$ or the projected one $\mathcal{P}(\bs{\hat{\tau}};\bs\Sigma).$

\begin{figure}[htb]
\begin{center}
\resizebox{8cm}{!}{
\includegraphics{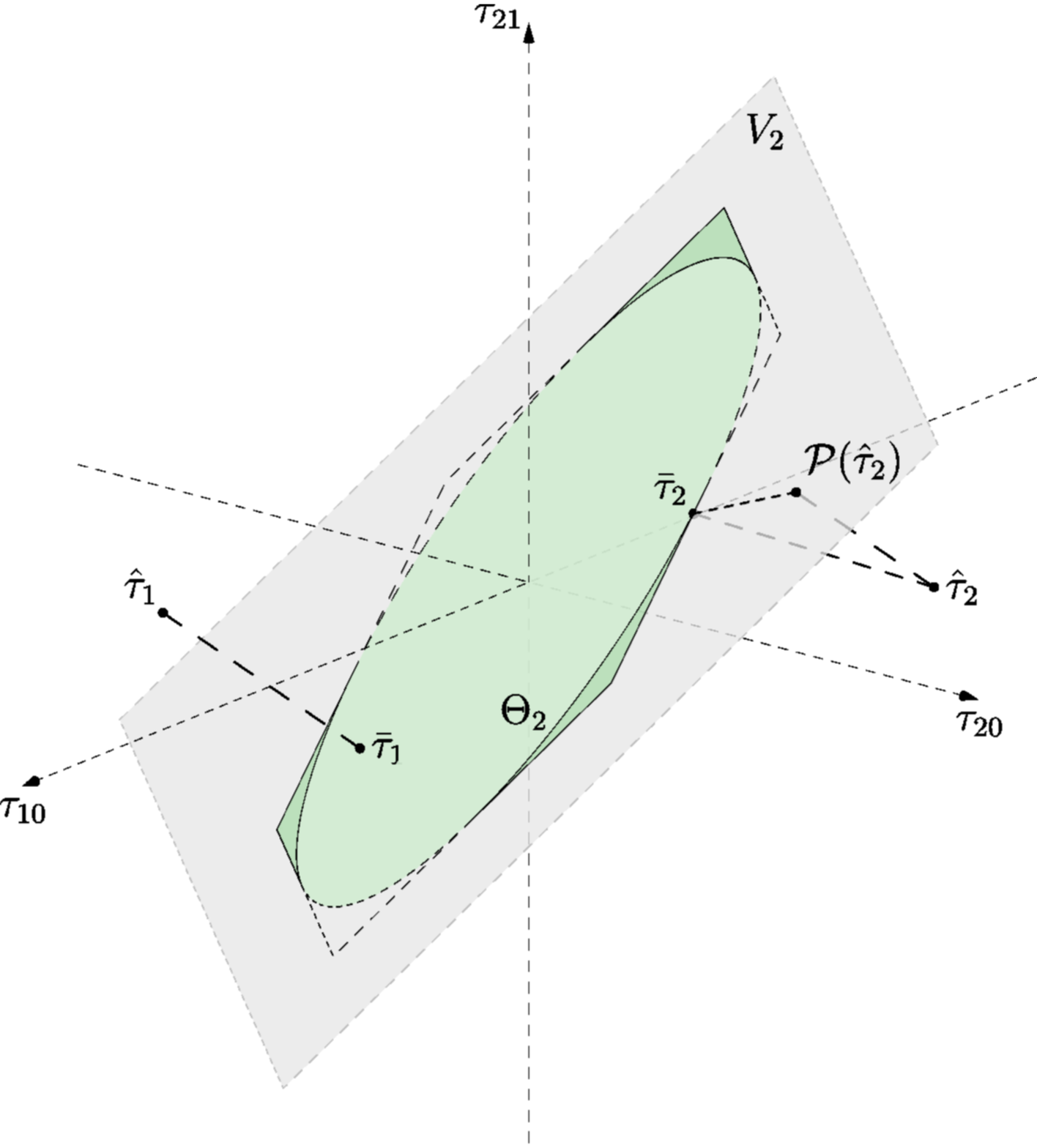}}
\caption{\label{fig:Denoising3D}
Denoising in the TDOA space, for the case of planar localization with $n=2$. A generic set of noisy TDOAs $\bs{\hat{\tau}}$ do not lie on $\Theta_2.$ The ML estimation finds the closest point $\bs{\bar{\tau}}_{\text{\tiny ML}}\in\Theta_2$ to $\bs{\hat{\tau}}.$ The solution of the relaxed denoisig problem is $\mathcal{P}(\bs{\hat{\tau}},\bs\Sigma)\in V_2.$ For the point $\bs{\hat{\tau}}_1$ it lies on $\Theta_2,$ then the projection $\mathcal{P}(\bs{\hat{\tau}}_1,\bs\Sigma)$ coincides with the ML solution $\bs{\bar{\tau}}_{1}$. For the point $\bs{\hat{\tau}}_2$ this is not true, thus $\mathcal{P}(\bs{\hat{\tau}}_1,\bs\Sigma)\neq \bs{\bar{\tau}}_{2}$. In any case, the projection gives a better estimate of the noiseless measurements $\bs{{\tau}},$ closest to $\bs{\bar{\tau}}_{\text{\tiny ML}}$ with respect to the original data point $\bs{\hat{\tau}}.$}
\end{center}
\end{figure}

We explicitly show this fact in Figure \ref{fig:Denoising3D}, for the case of planar localization with $n=2$. In this case one has two different situations, exemplified for the measurement points $\bs{\hat{\tau}}_1$ and $\bs{\hat{\tau}}_2$, respectively. In the first case, the orthogonal projection $\mathcal{P}(\bs{\hat{\tau}}_1;\bs\Sigma)\in\Theta_2$, thus it coincides with its ML estimate $\bs{\bar{\tau}}_{\text{\tiny ML},1}$. Differently, for the second point one has $\mathcal{P}(\bs{\hat{\tau}}_2;\bs\Sigma)\notin\Theta_2$. In this case, finding the corresponding ML solution $\bs{\bar{\tau}}_{\text{\tiny ML},2}$ corresponds to finding the closest point to $\mathcal{P}(\bs{\hat{\tau}}_2;\bs\Sigma)$ in $\Theta_2$, which implies solving a complicated minimization problem in $V_2$. Indeed, as we explained in Section \ref{sec:D2n2}, the feasible set $\Theta_2$ has a non trivial structure. In particular, its boundary is the union of six segments and three arcs of ellipse. If we define $\bs{\bar{\tau}}_{\text{\tiny ML},2}$ as the closest point to $\bs{\hat\tau}_2$ lying on one of these sets, then it is necessary to develop an ad hoc algorithm for finding it. Another complication is that $\Theta_2$ is not a closed set. This implies that if $\bs{\bar{\tau}}_{\text{\tiny ML},2}$ lies on the ellipse, then it does not correspond to a true source position because it is not part of $\Theta_2$. Conversely we should consider $\bs{\bar{\tau}}_{\text{\tiny ML},2}$ as the TDOAs associated to a source placed at infinity.

Thanks to the results and discussion contained in Section \ref{sec:generalTDOAspace}, we can generalize the previous analysis. The denoising problem \eqref{eq:MLE-TDOAspace2} can be subdivided into two subproblems.
\begin{enumerate}
\item
The easiest part is the projection on the linear subspace $V_n.$ We can call it the relaxed denoising problem
\begin{equation}\label{eq:MLE-TDOAspace_relaxed}
\mathcal{P}(\bs{\hat{\tau}};\bs\Sigma) =
\underset{\bt\in V_n}{\text{argmin}}\
\Vert\bs{\hat{\tau}}-\bt\Vert_{\bs\Sigma^{\unaryminus 1}}^2,
\end{equation}
in comparison to the complete denoising problem \eqref{eq:MLE-TDOAspace2}. Being the  search set in \eqref{eq:MLE-TDOAspace_relaxed} a linear subspace, the problem admits a closed solution.
\item
The hardest part is the projection of $\mathcal{P}(\bs{\hat{\tau}};\bs\Sigma)$ onto $\Theta_n:$ 
$$
\bs{\bar{\tau}}_{\text{\tiny ML}} = \underset{\bt\in\Theta_n}{\text{argmin}}\
\Vert\mathcal{P}(\bs{\hat{\tau}};\bs\Sigma) \middleminus \bt\Vert_{\bs\Sigma^{\unaryminus 1}}^2\,,
$$
From the discussion contained in Section \ref{sec:generalTDOAspace}, the difficulties are twofold: we have not the analytic description of $\Theta_n$ and, in any case, the feasible set is a complicated three dimensional semialgebraic variety embedded in the $n$-- dimensional linear subspace $V_n.$
\end{enumerate}

The previous observations are the original starting points for our interpretation and for the statistical justification of the relaxed denoising procedure. Indeed, observation 2 confirms the well known fact that solving ML estimation is a too difficult task, due to the non-linearity and non-convexity of the feasible set $\Theta_n$. However, observation 1 states that ML is composed by two parts with different difficulties. We can easily envision that the projection of the measured TDOAs onto the linear subspace $V_n$ leads to improvements in terms of localization accuracy, and this will be confirmed analytically in the forthcoming sections.

\subsection{The relaxed denoising algorithm and its statistical analysis}
\label{sec:relaxed_den_alg}
We can now proceed with a rigorous formulation of the intuition discussed above and with a precise definition of the relaxed denoising algorithm.

\subsubsection{The projection as a sufficient statistic}
\begin{theorem}\label{th:ss}
The orthogonal projection $\mathcal{P}(\bs{\hat{\tau}};\bs\Sigma)$ of $\bs{\hat{\tau}}\in\RR^q$ on $V_n,$ with respect to the scalar product $\langle\ ,\ \rangle_{\bs\Sigma^{-1}},$ is a sufficient statistic for the underlying parameter $\x.$
\end{theorem}
\noindent\emph{Proof:}
Let us start from
$$
\Vert\bs{\hat{\tau}} \middleminus \bs{\tau_n^*}(\mb{x})\Vert_{\bs\Sigma^{\unaryminus 1}}^2
=
\Vert\bs{\hat{\tau}} \middleminus \mathcal{P}(\bs{\hat{\tau}};\bs\Sigma)
\Vert_{\bs\Sigma^{\unaryminus 1}}^2+
\Vert\mathcal{P}(\bs{\hat{\tau}};\bs\Sigma) \middleminus \bs{\tau_n^*}(\mb{x})\Vert_{\bs\Sigma^{\unaryminus 1}}^2\,.
$$
Therefore, the probability density function \eqref{eq:TDOAprob} can be rewritten as
$$
p(\bs{\hat{\tau}};\bs{\tau_n^*}(\x),\bs\Sigma)=
\displaystyle
\frac{e^{-\frac{1}{2}\Vert\bs{\hat{\tau}}-\mathcal{P}(\bs{\hat{\tau}};\bs\Sigma)
\Vert_{\bs\Sigma^{\unaryminus 1}}^2}\,
e^{-\frac{1}{2}\Vert\mathcal{P}(\bs{\hat{\tau}};\bs\Sigma)-\bs{\tau_n^*}(\mb{x})\Vert_{\bs\Sigma^{\unaryminus 1}}^2}}
{\sqrt{(2\pi)^q|\bs\Sigma|}}\, .
$$
Then, the proof follows as a consequence of the Fisher-Neyman factorization theorem \citep{Casella1998}.
\hfill$\square$\vspace{1mm}\\
Theorem \ref{th:ss} states that the component of $\bs{\hat\tau}$ orthogonal to the linear subspace $V_n$ does not carry information on the source location and that we can remove it without corrupting the data.

\subsubsection{The algorithm}
For any $\bs{\hat{\tau}}\in\RR^q,$ the projection $\mathcal{P}(\bs{\hat{\tau}};\bs\Sigma)$ can be computed in closed form in two steps:
\begin{enumerate}[label=\Roman*., leftmargin=*, widest=iii]
\item
If we group all the equations \eqref{eq:lineq}, we end up with the homogeneous equation system
\begin{equation}
\label{eq:hom_TDOA_system}
\mathbf{C}\bs{\tau}=\mathbf{0}\,,
\end{equation}
where $\mathbf{C}$ is the $(q-n,q)$ matrix
$$
\mathbf{C}=\small\left(\begin{array}{cccccc:cccc}
-1 & 1 & 0 & \cdots & 0 & 0 & -1 & 0 & \cdots & 0\\
-1 & 0 & 1 & \cdots & 0 & 0 & 0 & -1 & \cdots & 0\\
\vdots & \vdots & \vdots &  & \vdots & \vdots &
\vdots & \vdots &  & \vdots\\
0 & 0 & 0 & \cdots & -1 & 1 & 0 & 0 & \cdots & -1
\end{array}
\right).
$$
The solution of this linear system is
$
V_n=\ker(\mathbf{C})\,.
$
In particular, we can find an orthonormal basis $\{\mb{v_1},\ldots,\mb{v_n}\}_{\bs\Sigma^{-1}}$ of $V_n,$ if necessary by using Gram-Schmidt algorithm defined with respect to the scalar product \eqref{eq:Fisherscalarprod}.
\item
The projection map $\mathcal{P}$ is defined on $\bs{\hat{\tau}}\in\RR^q$ as
\begin{equation}\label{eq:projmap}
\mathcal{P}(\bs{\hat{\tau}};\bs\Sigma)=
\langle\bs{\hat{\tau}},\mb{v_1}\rangle_{\bs\Sigma^{-1}}\;\mb{{v}_1} +\ldots+
\langle\bs{\hat{\tau}},\mb{v_n}\rangle_{\bs\Sigma^{-1}}\;\mb{{v}_n}.
\end{equation}
Let $\mb{e_{ji}},\ n\geq j>i\geq 1$ be the vectors in the standard basis $\mathcal{B}_q$ of $\RR^q.$ With respect to $\mathcal{B}_q$, the projection is represented by the $(q,q)$ matrix
\begin{equation}\label{eq:projVSmatrix}
\mathbf{P}=\left[\begin{array}{ccc}
\mathcal{P}(\mb{e}_{\mb{10}};\bs\Sigma) & \ldots & \mathcal{P}(\mb{e}_{\mb{n-1\,n}};\bs\Sigma)
\end{array}\right] \; .
\end{equation}
Consequently, the set of denoised TDOAs is obtained as
\begin{equation}\label{eq:projbymatrix}
\mathcal{P}(\bs{\hat{\tau}};\bs\Sigma) = \mathbf{P}\,\bs{\hat{\tau}} \; .
\end{equation}
\end{enumerate}

\subsubsection{The analysis of the noise reduction}\label{sec:analysisnoisered}
We can now compute the noise reduction on the TDOAs due to the relaxed denoising procedure defined above. Preliminarily, we prove the following Lemma.
\begin{lemma}\label{th:lemma}
	The covariance matrix $\bs\Sigma$ defines a Euclidean structure on $\RR^q$ and the matrix $\mathbf{P}^T$ represents an orthogonal projection with respect to the scalar product $\langle\ ,\ \rangle_{\bs\Sigma}.$
\end{lemma}
\noindent\emph{Proof:}
By construction $\bs\Sigma$ is symmetric and positive defined, therefore it defines a scalar product on $\RR^q.$ To prove that $\mb{P}^T$ represents an orthogonal projection with respect to this Euclidean structure, we have to show that $\mathbf{P}^T \mathbf{P}^T=\mathbf{P}^T$ and $\bs\Sigma \mathbf{P}^T=\mathbf{P}\bs\Sigma.$
On the other hand, we know that the matrix $\mathbf{P}$ represents an orthogonal projection with respect to $\langle\ ,\ \rangle_{{\bs\Sigma}^{-1}},$ hence $\mathbf{P} \mathbf{P}=\mathbf{P}$ and $\bs\Sigma^{-1} \mathbf{P}=\mathbf{P}^T\bs\Sigma^{-1}$. By using these identities, we have:
\begin{itemize}
	\item[$\bullet$]
	$\mathbf{P}^T \mathbf{P}^T=(\mathbf{P} \mathbf{P})^T=\mathbf{P}^T,$
	\item[$\bullet$]
	$\mathbf{P}\bs\Sigma=\bs\Sigma\bs\Sigma^{-1}\mathbf{P}\bs\Sigma=
	\bs\Sigma \mathbf{P}^T\bs\Sigma^{-1}\bs\Sigma=\bs\Sigma \mathbf{P}^T,$
\end{itemize}
which completes the proof.
\hfill$\square$\vspace{1mm}\\
A consequence of Lemma \ref{th:lemma} is that $\Vert\mb{v}\Vert_{\bs\Sigma}^2\geq\Vert \mathbf{P}^T\,\mb{v}\Vert_{\bs\Sigma}^2$ for any $\mb{v}\in\RR^q.$ This is useful for proving the next Theorem.
\begin{theorem}\label{th:sigmaproj}
	Let $\bs{\Sigma}$ be the covariance matrix of $\bs{\hat{\tau}}.$ Then:
	\begin{enumerate}
		\item
		the covariance matrix of $\mathcal{P}(\bs{\hat{\tau}};\bs\Sigma)$ is
		$\bs{\Sigma}'=\mathbf{P}\bs{\Sigma}\mathbf{P}^T;$
		\item
		we have $\bs{\Sigma}\succeq\bs{\Sigma}',$ i.e. $\bs{\Sigma}-\bs{\Sigma}'$ is positive semidefinite.
	\end{enumerate}
\end{theorem}
\noindent\emph{Proof:}
the first claim follows from the general transformation rule of the covariance matrix under linear mapping of $\bs{\hat{\tau}}$. Using this fact and Lemma \ref{th:lemma}, we have that $\mb{v}^T(\bs\Sigma-\bs\Sigma')\mb{v}=
\Vert\mb{v}\Vert_{\bs\Sigma}^2-\Vert \mathbf{P}^T\,\mb{v}\Vert_{\bs\Sigma}^2\geq 0$
for any $\mb{v}\in\RR^q,$ which completes the proof.
\hfill$\square$\vspace{1mm}\\
Theorem \ref{th:sigmaproj} states that the relaxed denoising procedure always reduces the noise on the TDOA dataset and it gives the way to quantify such reduction.

\subsection{Relation to state-of-the-art algorithms}
As already mentioned in the Introduction, the algorithm presented to solve the relaxed denoising problem is a different interpretation of other methods proposed for reducing the noise on TDOAs, exploiting data redundancy. In particular, authors in \citep{Schmidt1996} and \citep{Cheung2008} use the constraints in \eqref{eq:lineq} to relate the full set of $q$ TDOAs $\bs{\tau}$ to $n$ nonredundant TDOAs referred to a sensor. Selecting the first sensor as the reference one, the nonredundant set is $\bs{\tau}_\text{\tiny NR}=\left(\tau_{10}\,,\,\tau_{20}\,,\ldots\,,\tau_{n0}\right)^T$. The linear relation between the two sets is given by
$$
\bs{\tau}=\mathbf{G}\bs{\tau}_\text{\tiny NR}\;,\qquad \text{with} \; 
\mathbf{G}=
\begin{bmatrix}
\mathbf{I}_n \\
\mathbf{Y}
\end{bmatrix}\; \text{where}\;
\mathbf{Y}=\small\left(\begin{array}{cccccc}
-1 & 1 & 0 & \cdots & 0 & 0\\
-1 & 0 & 1 & \cdots & 0 & 0\\
\vdots & \vdots & \vdots &  & \vdots & \vdots\\
0 & 0 & 0 & \cdots & -1 & 1
\end{array}
\right)\;,
$$
and $\mathbf{I}_n$ is the identity matrix of order $n$.
Given the measured TDOAs $\bs{\hat{\tau}}$, in \citep{Schmidt1996} the nonredundant set is estimated in the least squares sense as 
$$
\bs{\hat{\tau}}_\text{\tiny NR} = (\mathbf{G}^T\mathbf{G})^{-1}\mathbf{G}^T\bs{\hat{\tau}}\;.
$$
In \citep{Cheung2008} this result is generalized accounting for the covariance structure of noise. To this purpose, the noise covariance matrix $\bs\Sigma$ is introduced in the weighted least squares solution
\begin{equation}
\label{eq:wls_cheung}
\bs{\hat{\tau}}_\text{\tiny NR} = (\mathbf{G}^T \bs\Sigma^{-1} \mathbf{G})^{-1}\mathbf{G}^T \bs\Sigma^{-1} \bs{\hat{\tau}}\;.
\end{equation}
Note that the two procedures coincide if $\bs\Sigma=\sigma^2\mb{I}.$ 

The nonredundant TDOAs $\bs{\hat{\tau}}_\text{\tiny NR}$ can be considered as the denoised measurements corresponding to the original set $\bs{\hat{\tau}}$. As a matter of fact, $\bs{\hat{\tau}}_\text{\tiny NR}$, computed as in \eqref{eq:wls_cheung}, coincides with the first $n$ components of the orthogonal projection $\mathcal{P}(\bs{\hat{\tau}};\bs\Sigma)$. At this respect, we can be more precise.
\begin{proposition}
Given the Euclidean structure on $\RR^q$ defined by $\bs{\Sigma}^{-1},$ the matrix $\mb{P}$ that represents the projection map $\mathcal{P},$ with respect to the canonical basis $\mathcal{B}_q,$ is
$$
\mb{P}=\mb{G}(\mathbf{G}^T \bs\Sigma^{-1} \mathbf{G})^{-1}\mathbf{G}^T \bs\Sigma^{-1}.
$$
It follows that $\bs{\hat{\tau}}_\text{\tiny NR}$ is a sufficient statistics for $\bs{\hat{\tau}}.$
\end{proposition}
\noindent\emph{Proof:}
First of all, we show that Im$(\mb{P})=V_n$.\footnote{With slight abuse of notation, in the proof we identify a matrix with the associated linear map defined through matrix multiplication.} We begin by observing that the matrix $\mb{G}^T \bs\Sigma^{-1} \mb{G}$ is invertible. Indeed, its kernel is given by the vectors $\x\in\RR^n$ such that $\mb{G}^T \bs\Sigma^{-1} \mb{G}\,\x=\mb{0}.$ In this case, we have
$$
\x^T\mb{G}^T \bs\Sigma^{-1} \mb{G}\,\x=\Vert\mb{G}\,\x\Vert_{\bs\Sigma^{\unaryminus 1}}^2=0,
$$
hence $\mb{G}\,\x=\mb{0}.$ Being rank$(\mb{G})=n,$ it follows that $\ker(\mb{G}^T \bs\Sigma^{-1} \mb{G})=\ker(\mb{G})=\{\mb{0}\}.$

From the previous point, we have
$$
\text{rank}((\mathbf{G}^T \bs\Sigma^{-1} \mathbf{G})^{-1}\mathbf{G}^T \bs\Sigma^{-1})=
\text{rank}(\mathbf{G}^T)=n
$$
and so Im($(\mathbf{G}^T \bs\Sigma^{-1} \mathbf{G})^{-1}\mathbf{G}^T \bs\Sigma^{-1})=\RR^n$.
It follows that Im$(\mb{P})=\text{Im}(\mb{G})=V_n.$

Now, it is straightforward to check that $\mb{P}^2=\mb{P}$ and $\bs{\Sigma}^{-1}\mb{P}=\mb{P}^T\bs\Sigma^{-1},$ thus $\mb{P}$ represents the projection map on Im$(\mb{P})=V_n$.

The last statement follows from the $1$--to--$1$ relation $\mathcal{P}(\bs{\hat\tau};\bs\Sigma)=\mb{G}\bs{\hat\tau}_\text{\tiny NR}.$
\hfill$\square$\vspace{1mm}

In the light of the above considerations, all our theorems and simulative results can be considered as further validations of state-of-the-art works in \citep{Schmidt1996} and \citep{Cheung2008}. In the following sections we will prove, both analytically and numerically, that any source localization algorithm benefits from denoised TDOAs. Moreover, we will see that this holds also when the full TDOA set is not completely available.
\section{Impact on source localization}\label{impact_on_loc}
To this point we have shown how it is possible to reduce the noise on a TDOA set $\bs{\hat{\tau}}$, thus approaching $\bs{\bar{\tau}}_{\text{\tiny ML}}$. The goal of this section is to prove that source location estimate benefits from the use of a denoised TDOA set. To this purpose, we first provide a solid theoretical analysis that demonstrates our claim. Then, we further investigate the effect on source localization by means of an extensive simulative campaign.

\subsection{Impact on sub-obtpimal localization algorithms}\label{sec:denoisingla}
The present analysis is mainly based on the results in Section \ref{sec:analysisnoisered}. In great generality, let us consider a given localization algorithm based on the minimization of a certain cost function $f(\x,\mb{c}),$ where $\mb{c}$ are the input TDOA data. Theorem \ref{th:sigmaproj} allows us to compare the accuracy of $\mb{\bar x}=\argmin f(\x,\bs{\hat{\tau}})$ and $\mb{\bar x}'=\argmin f(\x,\mathcal{P}(\bs{\hat{\tau}};\bs\Sigma)).$
Indeed, the first order error propagation formula given in \citep{Compagnoni2012} relates the covariance matrices $\bs\Sigma,\bs{\Sigma}'$ to $\bs{\Sigma}_{\mb{\bar x}},\bs{\Sigma}'_{\mb{\bar x}},$ respectively:
\begin{equation}
\bs{\Sigma}_{\mb{\bar x}}=\mathbf{A}(\x)\bs{\Sigma} \mathbf{A}(\x)^T
\quad\text{and}\quad
\bs{\Sigma}'_{\mb{\bar x}}=\mathbf{A}(\x)\bs{\Sigma}' \mathbf{A}(\x)^T,
\end{equation}
where $\mathbf{A}(\x)$ is defined by equation (26) from \citep{Compagnoni2012}. Then, by easily adapting the proof of Theorem \ref{th:sigmaproj}, we end up with
\begin{corollary}\label{th:sigmax}
	at first order approximation, $\bs{\Sigma}_{\mb{\bar x}}\succeq \bs{\Sigma}'_{\mb{\bar x}}.$
\end{corollary}
We remark that Corollary \ref{th:sigmax} is valid for every choice of the cost function $f(\x,\mb{c}).$ In particular, it includes the special cases of $f(\x,\mb{c})$ explicitly depending only on $n$ TDOAs, the ones relative to a reference microphone. In the following paragraph we will report concrete examples of these facts.

\subsection{Numerical examples}
We now analyze the effect of relaxed denoising\footnote{For the sake of compactness, throughout this paragraph we will use the word \textit{denoising} referring to \textit{relaxed denoising}.} on simulated data. Some results were reported in \citep{Schmidt1996,Cheung2008}. In particular, in \citep{Cheung2008} authors proved that the denoising procedure can be used to proficiently exploit the redundancy of the full set of TDOAs measured among all microphone pairs. Indeed, they numerically verified that ML source localization using the denoised TDOA set (denoted as optimum nonredundant set in \citep{Cheung2008}) meets the Cramer Rao Lower Bound (CRLB) \citep{Benesty2004} computed considering the full set of measurements. Note that the analytical proof of this fact is a direct consequence of Theorem \ref{th:sigmax}. 

Here we aim at providing a more comprehensive analysis, testing the effect of denoising on several source localization algorithms. In particular, we are interested in sub-optimal algorithms admitting an exact solution, i.e., for which the global optimum can be computed with no approximations. This feature is desirable in practice, as it prevents the risk of getting trapped in local minima, at the expense of obtaining a solution that does not meet the CRLB. To this end, we consider the following algorithms:
\begin{itemize}
	\item LS: unconstrained linear least squares estimator \citep{Smith1987}, admitting a closed-form solution. It uses the nonredundant TDOA set, measured considering a reference sensor;
	\item SRD-LS: squared-range-difference least squares estimator \citep{Beck2008}, a constrained version of the LS one, whose exact solution is computable efficiently. It is based on the nonredundant TDOA set;
	\item GS: Gillette-Silverman algorithm \citep{Gillette2008a}, an extension of the LS algorithm accommodating the case of multiple reference sensors. It can be applied to the full TDOA set.
\end{itemize}
Denoising was tested considering TDOAs corrupted by both Gaussian and non-Gaussian noise. The first case constitutes an ideal scenario, in which the noise model exactly meets the theoretical assumptions used for deriving the denoising procedure. The second case represents a non-ideal condition, useful to assess to what extent denoising is applicable when the assumptions are not obeyed.

\subsubsection{TDOAs corrupted by Gaussian noise}\label{sec:simulation_gaussian}
We simulated a compact cross array composed by $7$ microphones in positions $(0,0,0)^T$, $(\pm0.5, 0, 0)^T$, $(0, \pm0.5, 0)^T$, $(0, 0, \pm0.5)^T\, \mathrm{m}$. More than 500 sources are homogeneously distributed on a sphere centered at $(0,0,0)^T$, whose radius $d$ ranges from $0.5\, \mathrm{m}$ to $2.5\, \mathrm{m}$. The simulation setup is sketched in Figure \ref{fig:sim_setup}.
\begin{figure}[t]
	\centering
	\includegraphics[width=.6\columnwidth]{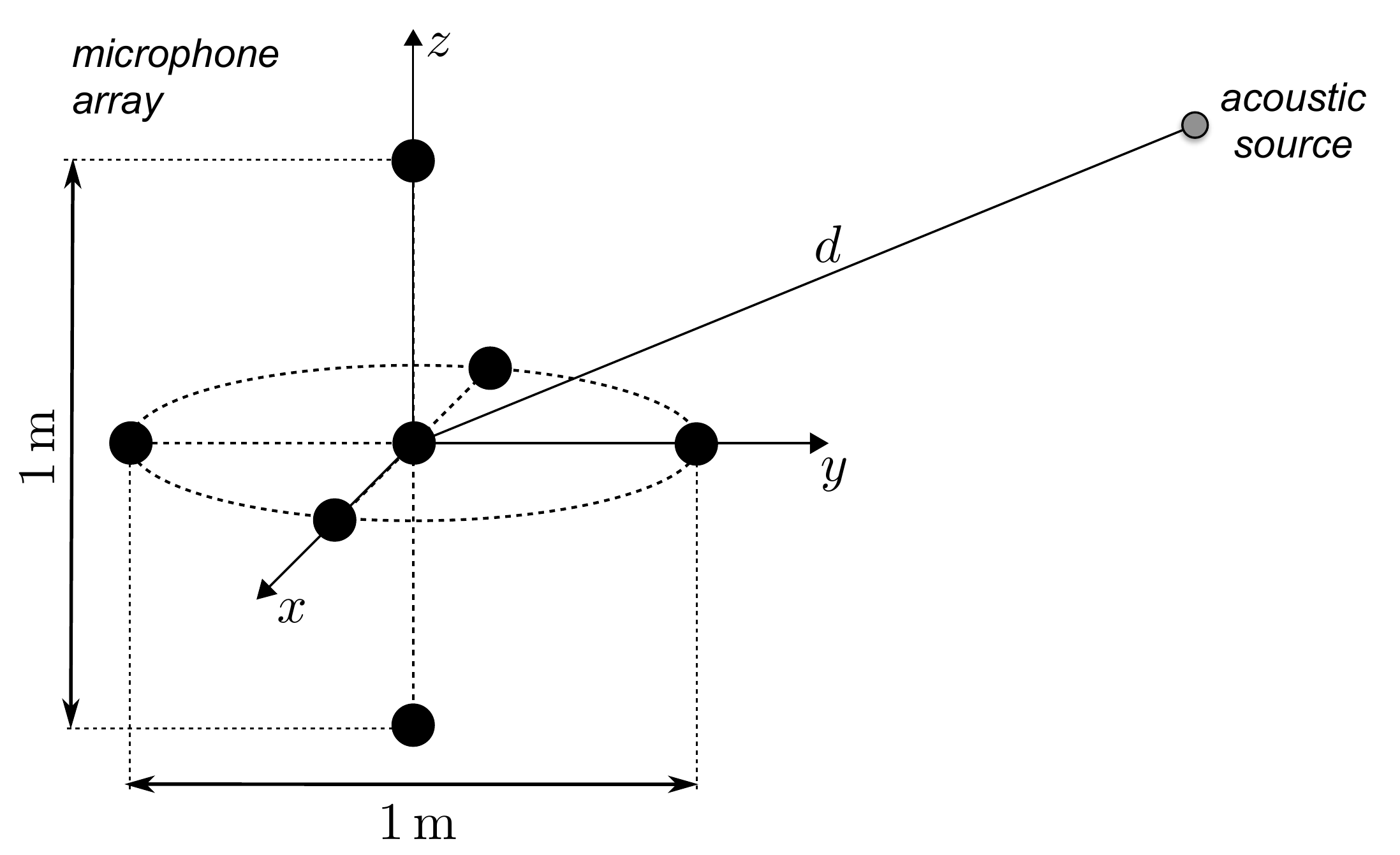}\label{fig:sim_setup}
	\caption{Simulation setup.}
	\label{fig:sim_setup}
\end{figure} 
For each source position, we computed the full set of $q=21$ theoretical TDOAs $\bs{\tau}$. We corrupted the vectors $\bs{\tau}$ with $I=5000$ realizations of i.i.d. zero-mean Gaussian noise with standard deviation $\sigma$, leading to the noisy TDOAs $\bs{\hat{\tau}}_i$. The covariance matrix thus resulted in $\mathbf{\Sigma}=\sigma^2 \mathbf{I}_q$, where $\mathbf{I}_q$ is the identity matrix of order $q$. Monte-Carlo simulations were carried out considering the range $\sigma \in [0.5\,\mathrm{cm}, 5\,\mathrm{cm}]$. The corresponding denoised TDOAs $\mathcal{P}(\bs{\hat{\tau}}_i;\bs\Sigma)$ were computed using \eqref{eq:projbymatrix}.

\begin{figure}[t]
	\centering
	\subfloat[]{\includegraphics[height=.33\columnwidth]{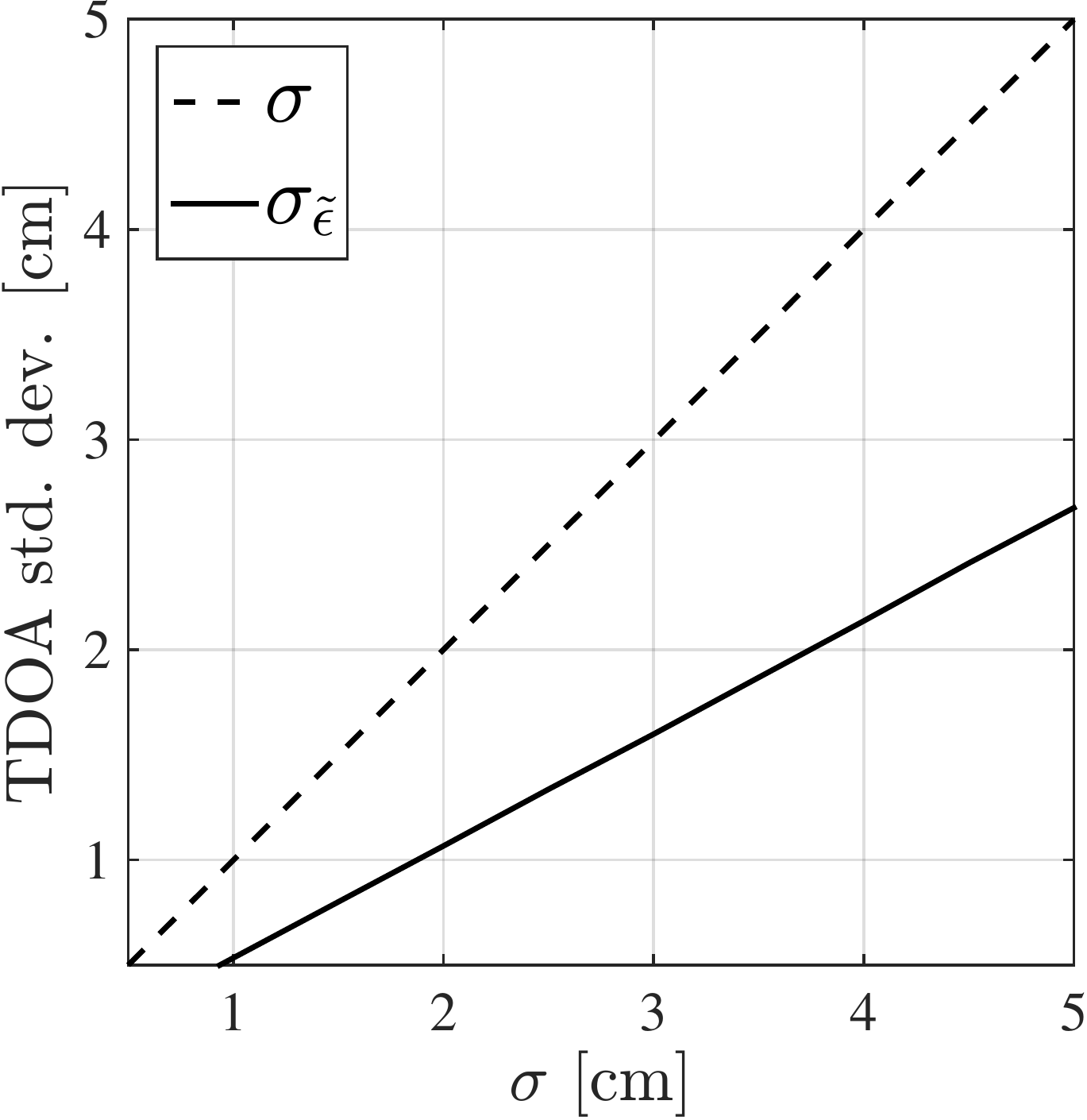}\label{fig:test1_TDOAerr_sigma}} \hfil
	\subfloat[]{\includegraphics[height=.325\columnwidth]{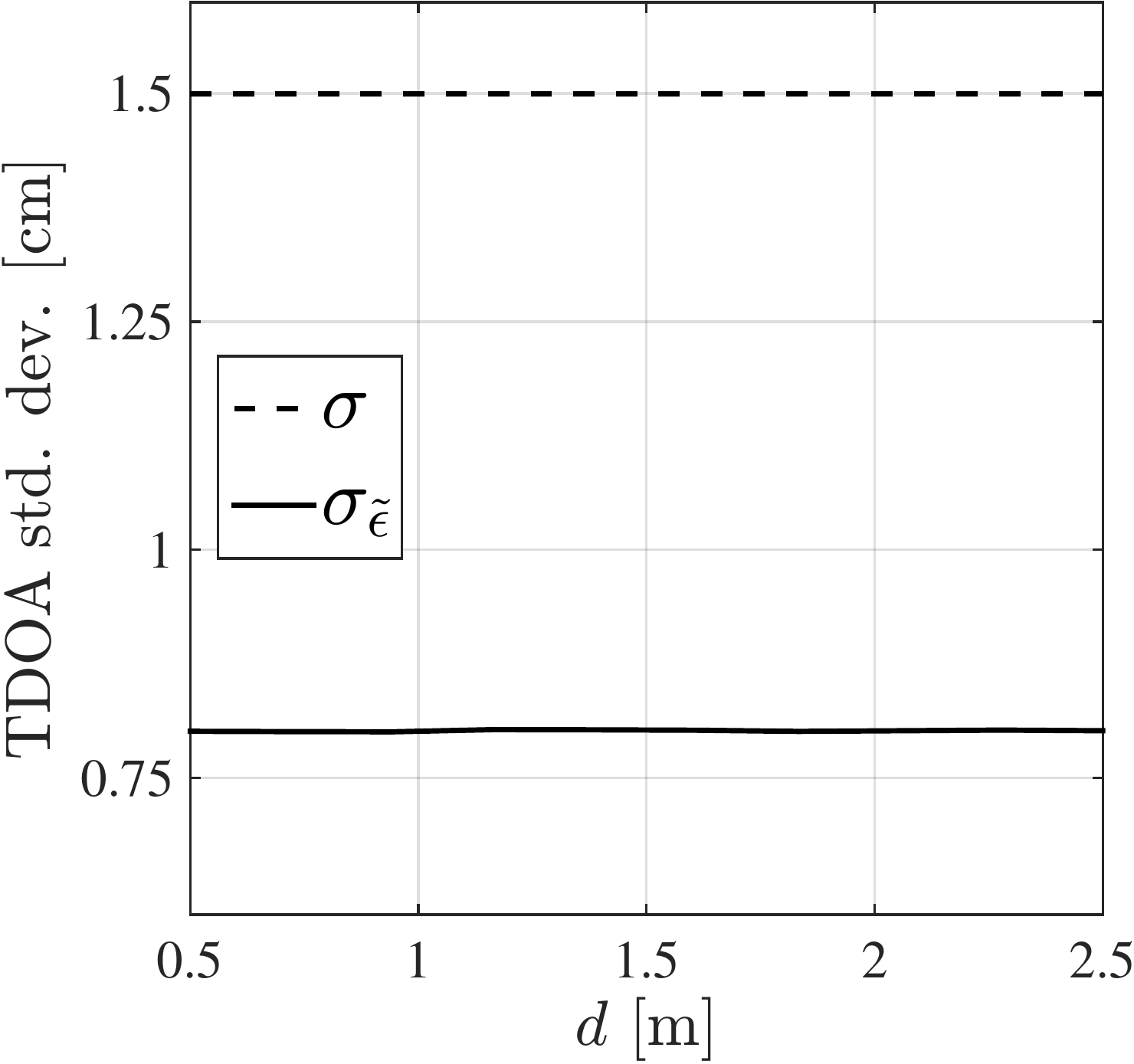}\label{fig:test1_TDOAerr_dist}}
	\caption{TDOA residual error before and after denoising, as a function of: injected noise $\sigma$ (a); source distance $d$ (b).}
	\label{fig:test1_TDOAerr}
\end{figure}

As a preliminary test, we computed the mean ${\mu}_{\bs{\tilde{\epsilon}}}$ and the standard deviation ${\sigma}_{\bs{\tilde{\epsilon}}}$ of the residual error $\bs{\epsilon}_i = \bs{\tilde{\tau}}_i - \bs{\tau}$ left on denoised TDOAs. We first observed that ${\mu}_{\bs{\tilde{\epsilon}}}$ is always negligible compared to ${\sigma}_{\bs{\tilde{\epsilon}}}$, meaning that the denoising procedure does not introduce relevant bias on TDOAs. Figure~\ref{fig:test1_TDOAerr_sigma} shows the standard deviation of denoised TDOAs $\sigma_{\tilde{\epsilon}}$ as a function of $\sigma$, averaged for sources located at a fixed distance $d=1.5\, \mathrm{m}$. Similarly, Figure~\ref{fig:test1_TDOAerr_dist} shows $\sigma_{\tilde{\epsilon}}$ for different distance values, fixing $\sigma=1.5\,\mathrm{cm}$. In both the cases, we notice that $\sigma_{\tilde{\epsilon}}$ (solid line) is always significantly below the value of the standard deviation of the injected noise $\sigma$ (dashed line), thus confirming the effectiveness of denoising. We also observe that $\sigma_{\tilde{\epsilon}} \approx \frac{1}{2} \sigma$, independently from the source position. 

\begin{figure}[t]
	\centering
	\subfloat[]{\includegraphics[height=.33\columnwidth]{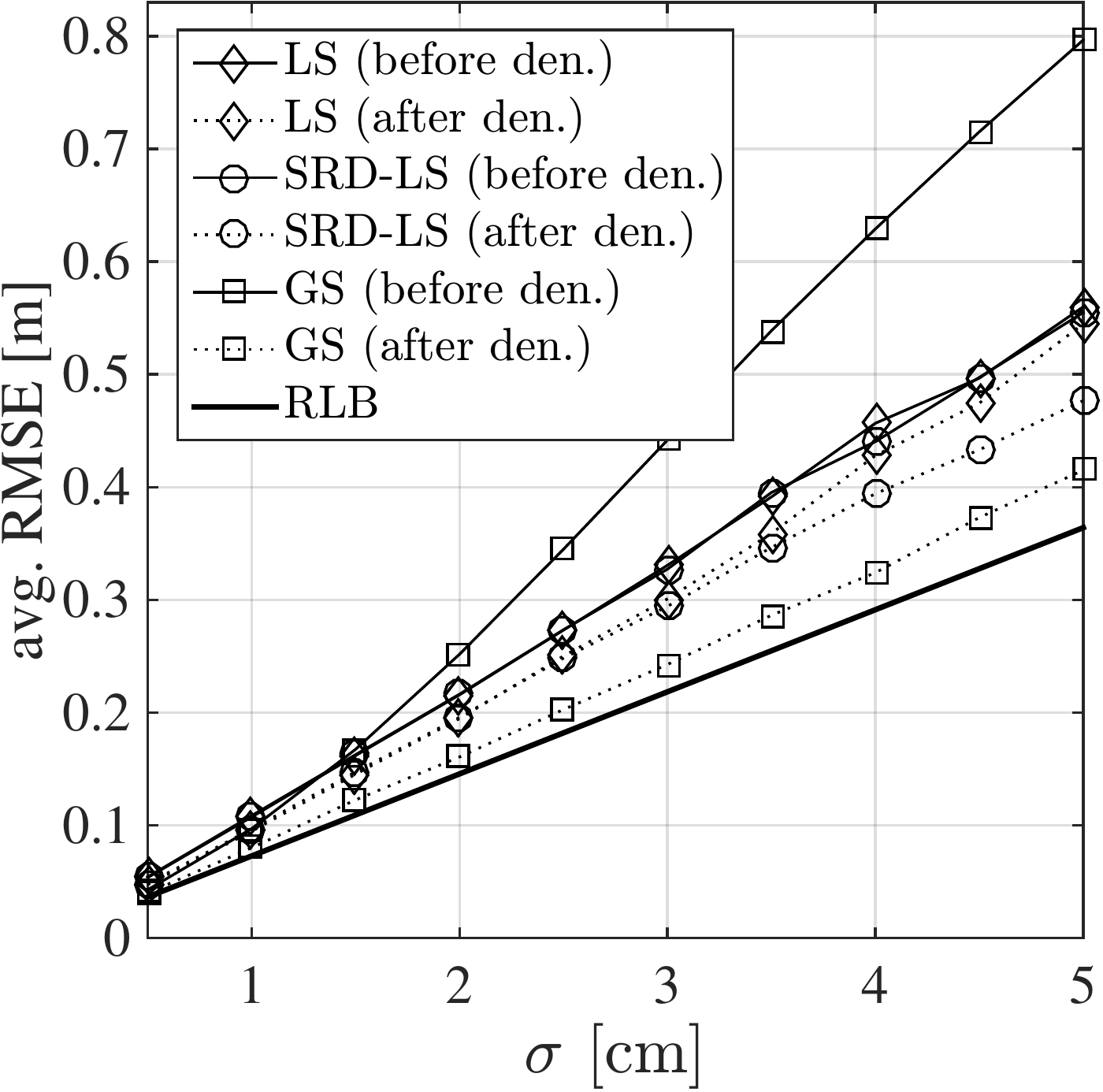}\label{fig:test1_RMSE_sigma}} \hfil
	\subfloat[]{\includegraphics[height=.335\columnwidth]{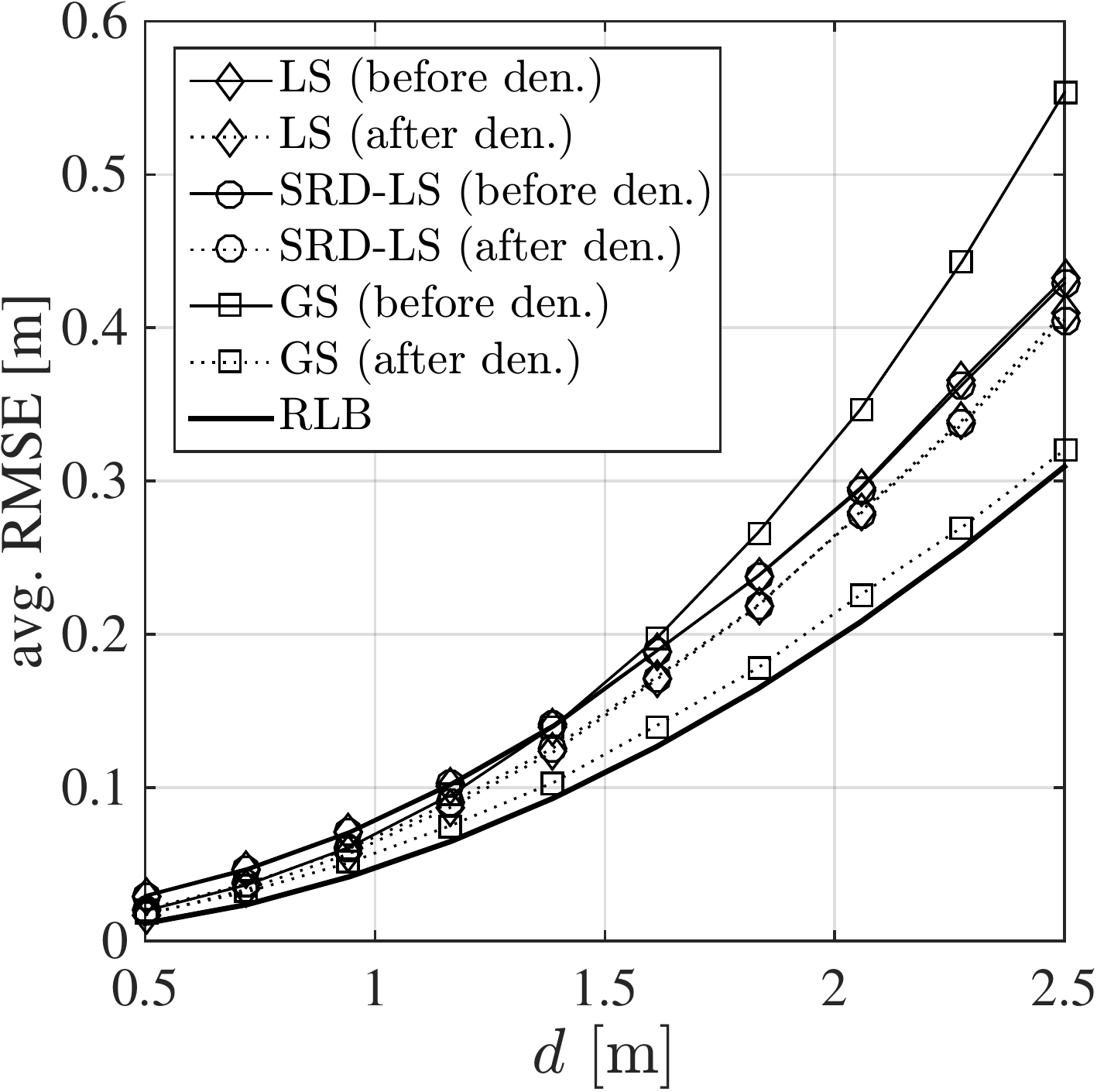}\label{fig:test1_RMSE_dist}}
	\caption{Localization accuracy before and after denoising, as a function of: injected noise $\sigma$(a); source distance $d$ (b).}
	\label{fig:test1_RMSE}
\end{figure}

The localization performance was then evaluated in terms of Root Mean Square Error (RMSE), computed as
\begin{equation}
\text{RMSE}(\x) = \sqrt{ \frac{1}{I} \sum_{i=1}^I \left\Vert \mathbf{\tilde{x}}_i - \mathbf{x} \right\Vert_2^2 } \; ,
\end{equation}
where $\mathbf{x}$ is the nominal source position and $\mathbf{\tilde{x}}_i$ is its estimate at the $i$th Monte-Carlo run. The three algorithms were fed with both the measured TDOAs $\bs{\hat\tau}_i$ and the denoised TDOAs $\mathcal{P}(\bs{\hat{\tau}_i};\bs\Sigma)$. In particular, for LS and SRD-LS we considered only the $n=6$ TDOAs measured with respect to the first sensor, selected as the reference one. For GS, we considered the full set of $q$ TDOAs, before and after denoising. Results are reported in Figure \ref{fig:test1_RMSE}.
As done before, we report results as a function of $\sigma$ when $d=1.5\, \mathrm{m}$ (Figure~\ref{fig:test1_RMSE_sigma}); and varying the distance $d$ when $\sigma=1.5\,\mathrm{cm}$ (Figure~\ref{fig:test1_RMSE_dist}). For each tested algorithm, the figures show the average RMSE achieved before and after the denoising of the measured TDOAs. For the sake of comparison, we also report the RMSE Lower Bound (RLB) implied by the CRLB. As expected, all the algorithms exhibit improved localization accuracy when denoised TDOAs are used. It is interesting to observe the different behavior of the algorithms using only the nonredundant TDOAs (LS and SRD-LS) and that using the full TDOA set (GS). Indeed, denoising produces a moderate effect on LS and SRD-LS, while the benefit is impressive for GS. Despite using the full TDOA set, before denoising GS exhibits very poor performances compared to LS and SRD-LS, especially for distant sources and at high amount of injected noise on TDOAs.
This behavior is not completely unexpected, as GS extends the LS approach, which is known to suffer from non negligible bias \citep{Benesty2004}. We observed that, using the full TDOA set, the bias increases significantly, even if the variance of estimation is reduced due to the availability of more measurements. We preliminarily noticed that the bias is greatly reduced when GS is fed with the denoised TDOA set. Roughly speaking, this means that denoising allows GS to effectively exploit the data redundancy. This positively impacts on source localization with GS after denoising, producing an RMSE that approaches the RLB. On the other hand, the redundancy is only partially exploited by LS and SRD-LS, as they rely on the nonredundant denoised TDOA set.

\subsubsection{TDOAs corrupted with non-Gaussian noise}
\begin{figure}[t]
	\centering
	\subfloat[]{\includegraphics[height=.33\columnwidth]{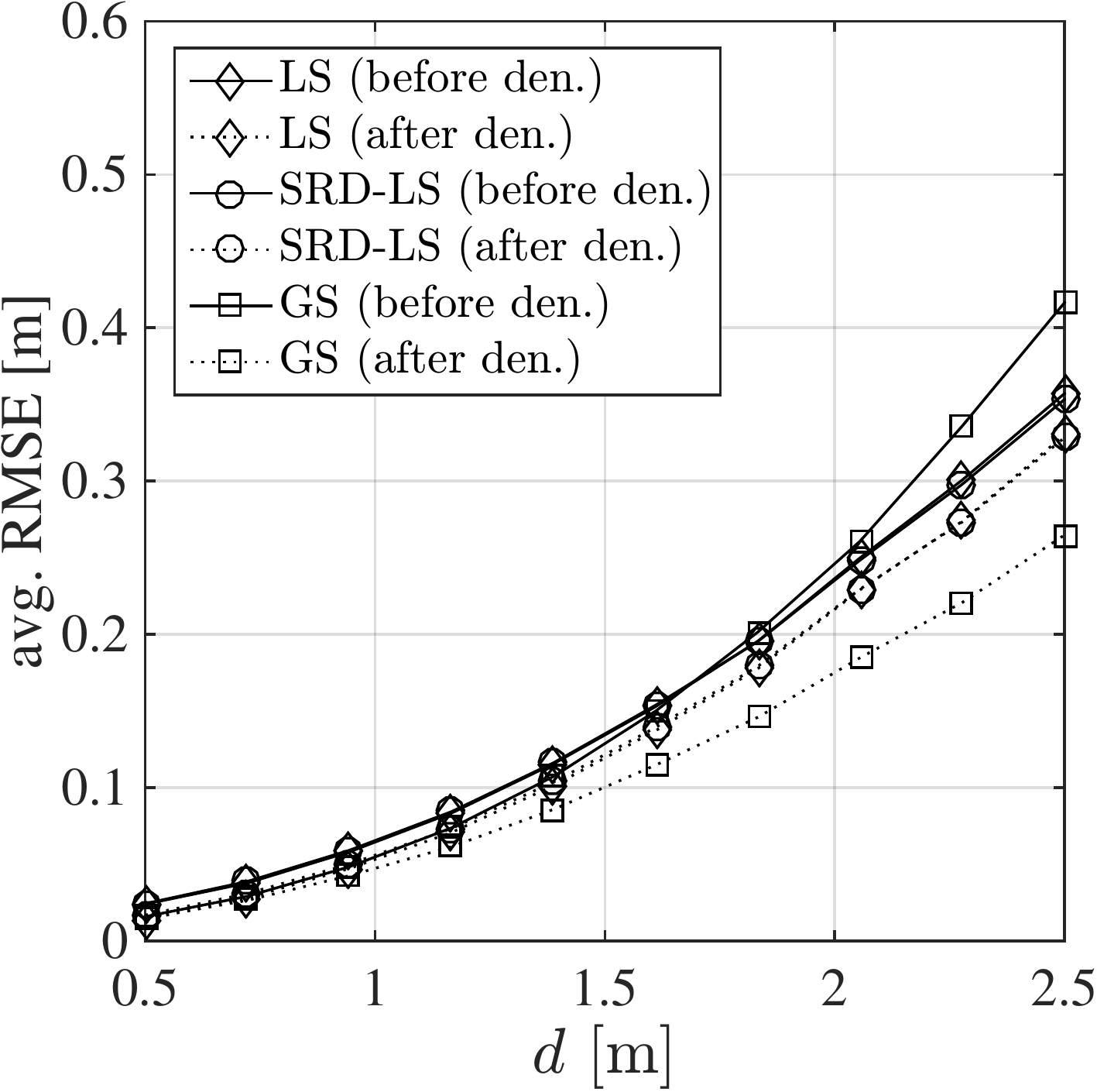}\label{fig:RMSE_uniform}} \hfil
	\subfloat[]{\includegraphics[height=.33\columnwidth]{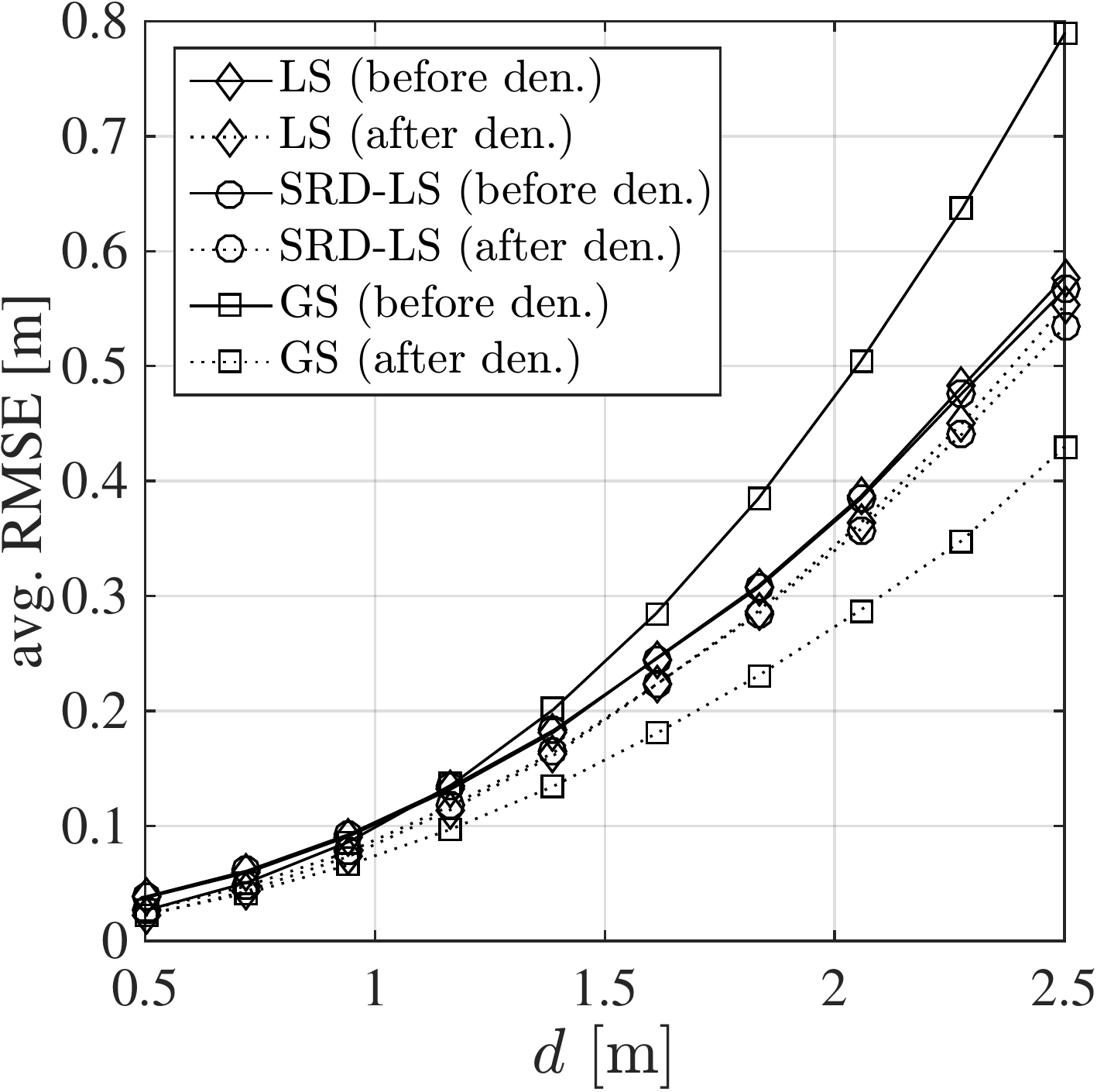}\label{fig:RMSE_uniform_and_gaussian}} \hfil
	\subfloat[]{\includegraphics[height=.33\columnwidth]{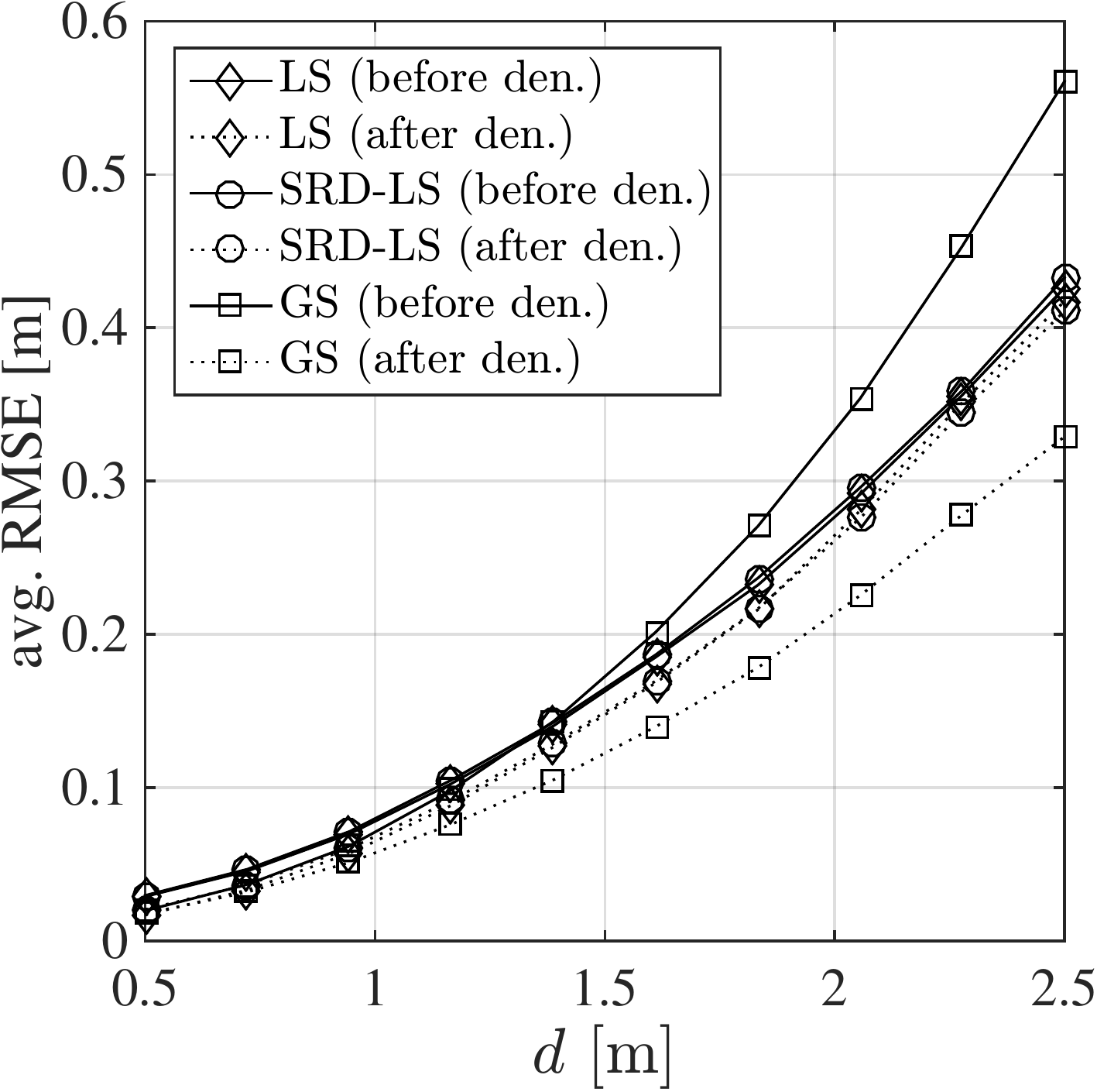}\label{fig:RMSE_laplacian}}
	\caption{Localization accuracy before and after denoising, as a function of distance $d$, for different TDOA noise models: uniformly distributed noise (a), mixture of uniform and Gaussian noise (b), Laplacian noise (c).}
	\label{fig:RMSE_noiseModels}
\end{figure}

Considering the simulation setup described in the previous paragraph, we repeated all the tests injecting different types of non-Gaussian noise in the ideal TDOAs. In particular, we focus on the following noise models:
\begin{itemize}
	\item i.i.d. zero-mean uniform noise in the range $\left[-\frac{c}{2f_s},\frac{c}{2f_s}\right]$, which mimics the sampling error caused by estimating the TDOA from the a Generalized Cross-Correlation (GCC) function \citep{Knapp1976} sampled at $f_s=8\,\textrm{kHz}$; $c=343\,\mathrm{m/s}$ is the speed of sound;
	\item a mixture of i.i.d zero-mean uniform and Gaussian noise, the former distributed in the range $\left[-\frac{c}{2f_s},\frac{c}{2f_s}\right]$; the latter with standard deviation $\sigma = 1.5\,\mathrm{cm}$. This mixture of models is suggested in \citep{Benesty2004};
	\item i.i.d. zero-mean Laplacian noise with standard deviation $\sigma = 1.5\,\mathrm{cm}$.
\end{itemize} 
The localization accuracy of the three algorithms exhibits the same trend obtained for the Gaussian noise model. 
For reasons of space, here we limit to report the average RMSE as a function of the source distance $d$. Figures \ref{fig:RMSE_uniform}, \ref{fig:RMSE_uniform_and_gaussian} and \ref{fig:RMSE_laplacian} show the results for the three considered noise distributions, respectively. 
Even if the noise models differ from the Gaussian assumption underlying the denoising theory, all the algorithms improve their accuracy using denoised TDOAs. As before, this is especially true for GS, which exhibit the most noticeable improvement. This suggests that denoising can be applied when the TDOA error does not strictly meet the Gaussian assumption.

\section{Dealing with an incomplete set of TDOAs}\label{sec:denoisinglTDOA}
There exist many scenarios where not the whole set of TDOAs is available. For example, when computational cost is an issue, the computation of all the possible TDOAs is not feasible. In the following, we adapt our previous analysis in order to handle relaxed denoising also in such situations.

Let us assume that the TDOAs $\{\tau_{j_1 i_1},\ldots,\tau_{j_s i_s}\},\ 0\leq s \leq q,$ are not available and let $S=\{(j_1,i_1),\ldots,(j_s,i_s)\}$ be the corresponding set of indices.
In this setting, the proper TDOA map is 
\begin{equation}\label{eq:TDOAmapred}
\begin{array}{cccl}
\bs{\tau_{n,S}^*}: & \RR^3 & \longrightarrow & \RR^{q-s}\\
& \x & \longmapsto & \bs{\tau_{n,S}^*}(\x)
\end{array}\,,
\end{equation}
where
$$
\bs{\tau_{n,S}^*}(\x)=(\tau_{10}(\x),\unarydots,\widehat{\tau_{j_1i_1}(\x)},\unarydots,\widehat{\tau_{j_si_s}(\x)},\unarydots,\tau_{n\,n-1}(\x))^{T}
$$
and $\widehat{\tau_{j_1i_1}(\x)}$ means that the item is missing. As before, we define the TDOA space as the target set $\RR^{q-s}$ of $\bs{\tau_{n,S}^*}$ and the image $\text{Im}(\bs{\tau_{n,S}^*})$ as $\Theta_{n,S}$. 
Clearly, the TDOA map $\bs{\tau_{n,S}^*}$ is strictly related to $\bs{\tau_n^*}.$
Indeed, let us consider the projection $\mathit{p}_S:\RR^q\rightarrow\RR^{q-s}$ that takes care of forgetting the $s$ coordinates corresponding to the indices in $S$.
Then, one has $\bs{\tau_{n,S}^*}=\mathit{p}_S\circ\bs{\tau_{n}^*}$ and $\Theta_{n,S}=\mathit{p}_S(\Theta_n)$, where the symbol $\circ$ denotes the function composition operator.

In the presence of measurement errors, we assume that the TDOAs are described by the statistical model
\begin{equation}\label{eq:TDOAstatmod_2}
\bs{\hat{\tau}_{n,S}^*}(\x)=\bs{\tau_{n,S}^*}(\x)+\bs{\varepsilon_S},\quad \text{where}\quad \bs{\varepsilon_S}\sim N(\mb{0},\bs{\Sigma_S}).
\end{equation}
As above, the Fisher matrix $\bs{\Sigma_S}^{-1}$ defines a Euclidean structure on the TDOA space $\RR^{q-s}$ and the same discussion made in Section \ref{sec:TS} holds in this situation.

The definition and the analysis of the relaxed denoising procedure are very similar to the ones made in Section \ref{sec:relaxed_den_alg}. First of all, we adapt Theorem \ref{th:ss}.
\begin{theorem}\label{th:redss}
	Let us take $n+1$ microphones at $\m{0},\ldots,\m{n}$ in $\RR^3$, where $n\geq 2.$ Then $\Theta_{n,S}$ is a subset of the linear subspace $V_S=\mathit{p}_S(V_n)\subset\RR^{q-s}.$ Moreover, the orthogonal projection $\mathcal{P}_{S}(\bs{\hat\tau_S};\bs\Sigma_{\bs{S}})$ of $\bs{\hat\tau_S}\in\RR^{q-s}$ on $V_S,$ with respect to $\langle\ ,\ \rangle_{\bs\Sigma_{\bs{S}}^{-1}},$ is a sufficient statistic for the underlying parameter $\x.$
\end{theorem}
\noindent\emph{Proof:}
$\mathit{p}_S$ is a linear map, therefore $V_S$ is a linear subspace of $\RR^{q-s}.$ The other claims follow in the same way of Proposition \ref{prop:Vn} and Theorem \ref{th:ss}.
\hfill$\square$\vspace{1mm}

From Theorem \ref{th:redss} it follows that $\dim(V_S)\leq\dim(V_n)=n$, where $\dim$ denotes the dimension of a vector space. It is not difficult to prove that the equality holds if, and only if, the set of available TDOAs contains $n$ independent measures, for example the $n$ TDOAs calculated with respect to a reference microphone. In this case, the map $\mathit{p}_S$ is a bijection between $V$ and $V_S.$ This means that it is possible to obtain the full set of $q$ denoised TDOAs as $\mathit{p}_S^{-1}(\mathcal{P}_S(\bs{\hat{\tau}_S};\bs\Sigma_{\bs{S}}))\in V.$ Concretely, one has to use the linear equations
\begin{equation}\label{eq:lineqred}
-\tau_{ik}+\tau_{jk}-\tau_{ji}=0,\qquad i\neq j\neq k,
\end{equation}
to calculate the missing TDOAs with indices in $S.$ It is important to remark that this operation does not increase the noise on the dataset, as indeed it would happen if we apply the same procedure on the original data by calculating $\mathit{p}_S^{-1}(\bs{\hat{\tau}_S}).$

\subsection{Denoising algorithm}
In order to explicitly construct the projection map $\mathcal{P}_{S},$ let us take the $(q-s,q)$ matrix $\mb{I_S}$ defined by removing the $s$ rows corresponding to the indexes in $S$ from the $(q,q)$ identity matrix. It can be easily proven that $\mb{I_S}$ represents the map $\mathit{p}_S$ with respect to the standard basis $\mathcal{B}_q$ and $\mathcal{B}_{q-s}$ of $\RR^q$ and $\RR^{q-s},$ respectively.
Then, given a generic basis  $\{\mb{v_1},\ldots,\mb{v_n}\}$ of $\ker(\mathbf{C})$, from Theorem \ref{th:redss} follows that the set $\{\mb{I_S\, v_1},\ldots,\mb{I_S\, v_n}\}$ spans $V_S.$ After reducing it to an independent set of vectors and subsequently applying the Gram-Schmidt algorithm, we can finally find an orthonormal basis of $V_S$ with respect to the scalar product $\langle\ ,\ \rangle_{\bs\Sigma_{\bs{S}}^{-1}}.$

From this point on, one can proceed exactly as done in the previous sections. In particular, the map $\mathcal{P}_S$ is defined by the analogous of the formula \eqref{eq:projmap} and it is represented, with respect to $\mathcal{B}_{q-s}$, by a $(q-s,q-s)$ matrix $\mb{P_S}.$ Then, the denoised TDOAs are
\begin{equation}\label{eq:projbymatrixred}
\mathcal{P}_S(\bs{\hat{\tau}_S};\bs\Sigma_{\bs{S}}) = \mathbf{P_S}\,\bs{\hat{\tau}_S}\; .
\end{equation}

\subsection{Impact on source localization}
We summarize the main facts on the denoising procedure in the following theorem.
\begin{theorem}\label{th:sigmaprojred}
	let $\bs{\Sigma_S}$ be the covariance matrix of $\bs{\hat{\tau}_S}$ and $f(\x,\mb{c})$ be any given cost function, where $\mb{c}$ are the input TDOA data. Then:
	\begin{enumerate}
		\item
		the covariance matrix of $\mathcal{P}_S(\bs{\hat{\tau}_S};\bs\Sigma_{\bs{S}})$ is
		$\bs{\Sigma_S}'=\mathbf{P_S}\bs{\Sigma_S}\mathbf{P_S}^T;$
		\item
		$\bs{\Sigma_S}\succeq\bs{\Sigma_S}';$
		\item
		at first order approximation, the covariance matrices $\bs{\Sigma}_{\bs S},\mb{\bar x}$ and $\bs{\Sigma}'_{\bs{S}},\mb{\bar x}$ of the estimators $\mb{\bar x}=\argmin f(\x,\bs{\hat{\tau}_S})$ and $\mb{\bar x}'=\argmin f(\x,\mathcal{P}_S(\bs{\hat{\tau}_S};\bs\Sigma_{\bs{S}})),$ respectively, satisfy
		$\bs{\Sigma}_{\bs{S},\mb{\bar x}}\succeq \bs{\Sigma}'_{\bs{S},\mb{\bar x}}.$ 
	\end{enumerate}
\end{theorem}
\noindent\emph{Proof:}
The proof is similar to the ones of Theorem \ref{th:sigmaproj} and Corollary \ref{th:sigmax}.
\hfill$\square$\vspace{1mm}

\subsection{Numerical examples}
\begin{figure}[t]
	\centering
	\subfloat[]{\includegraphics[width=.33\columnwidth]{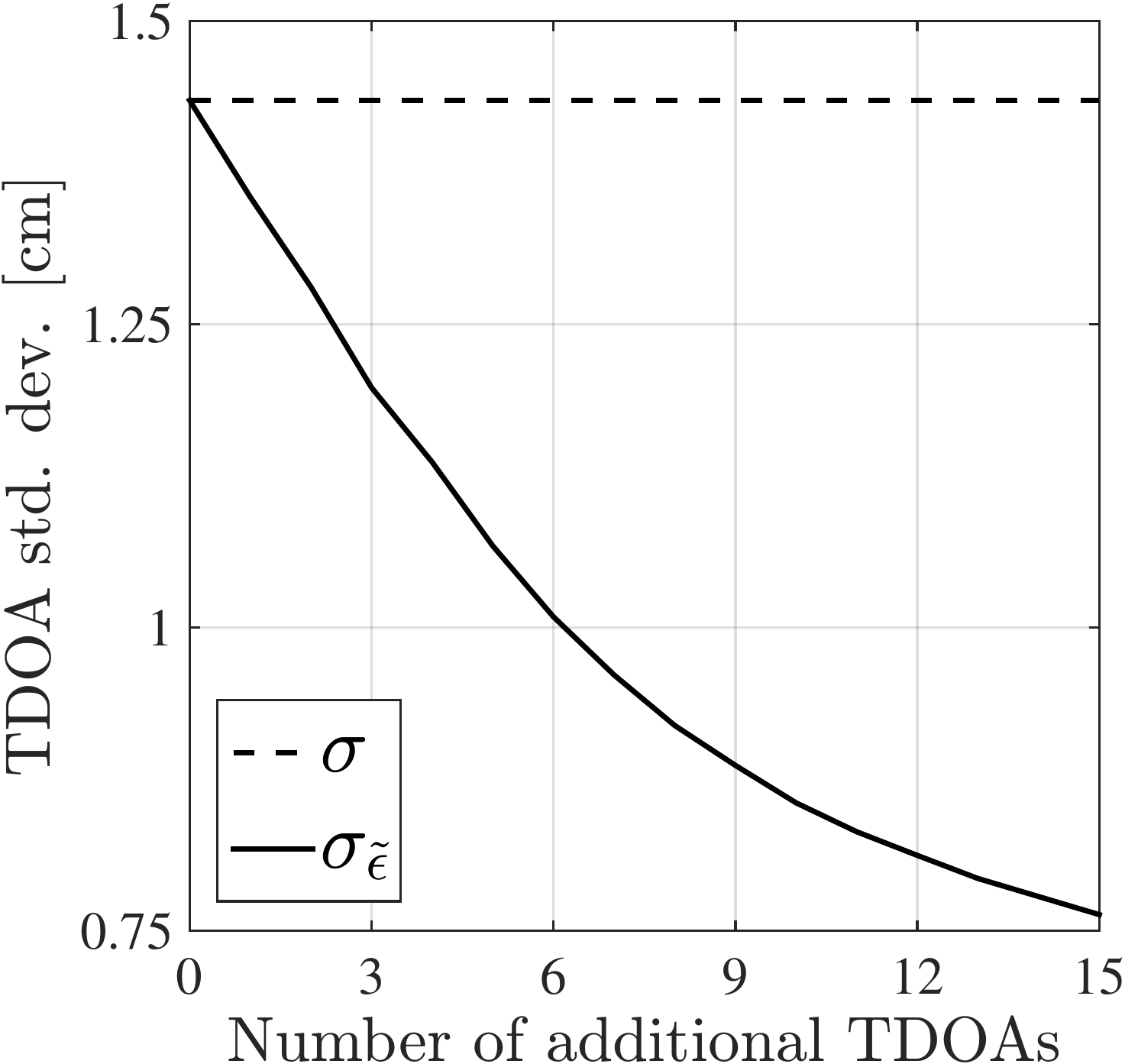}\label{fig:test3_TDOAerr}} \hfil
	\subfloat[]{\includegraphics[width=.33\columnwidth]{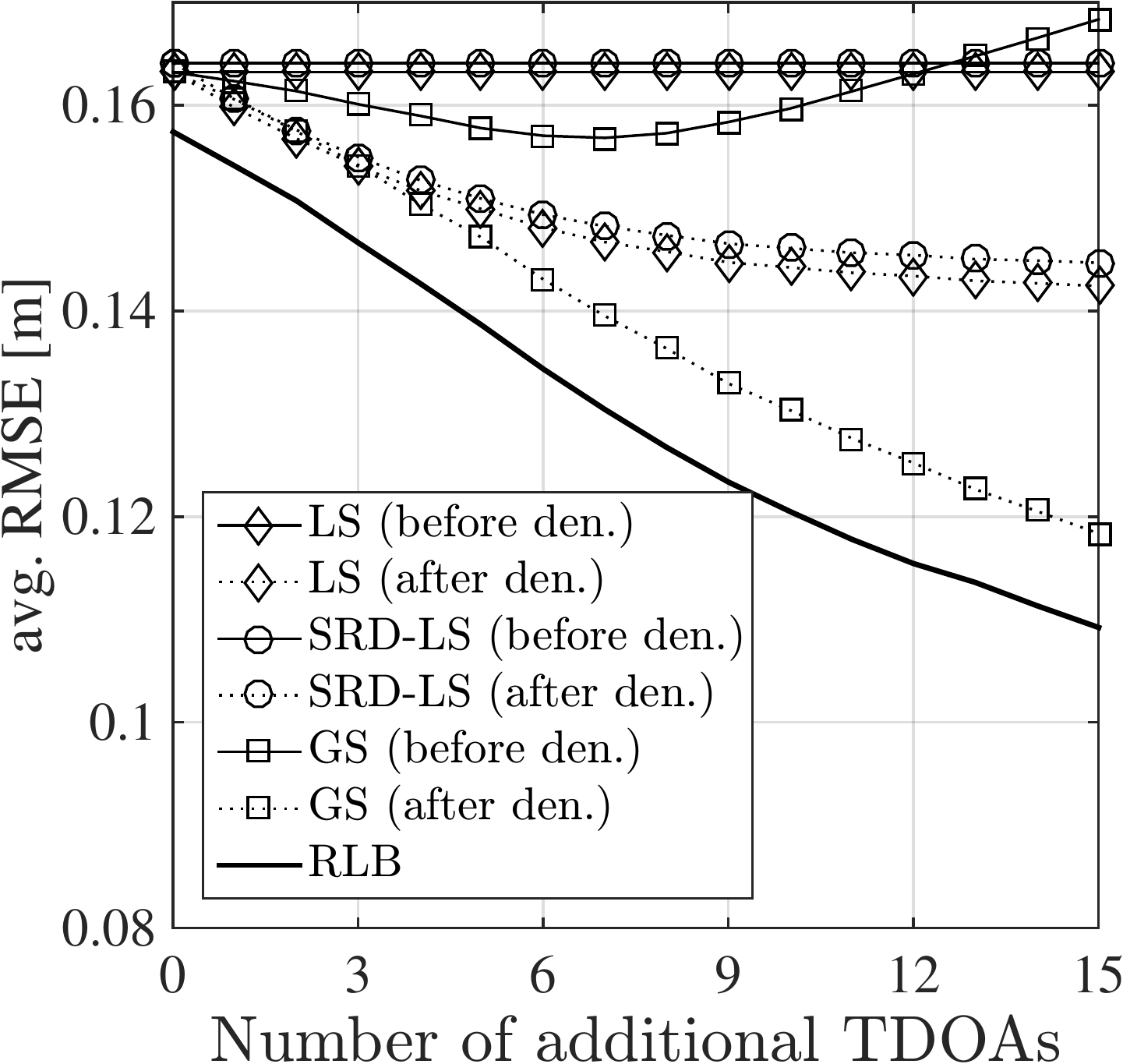}\label{fig:test3_locRMSE}}
	\caption{TDOA residual error before and after denoising (a)  and localization accuracy (b). Both plots are function of the number of additional TDOAs}
	\label{fig:test3}
\end{figure}

In this paragraph we show some numerical examples, devoted to investigate the effect of relaxed denoising when the full TDOA set is not entirely available. To this end, we refer again to the simulation setup described in Section \ref{sec:simulation_gaussian}. However, we now consider the availability of the $n$ TDOAs referred to the first microphone, along with $z \leq q - n$ additional TDOAs. Noisy TDOAs were obtained corrupting their nominal values with i.i.d. zero-mean Gaussian noise with standard deviation $\sigma=1.5\,\mathrm{cm}$. In this scenario, denoising was accomplished using \eqref{eq:projbymatrixred}. In particular, we built the vector $\bs{\hat{\tau}_S}$ including the $n+z$ available TDOAs, and we computed the projection matrix $\mathbf{P_S}$ accordingly. For this test we considered sources at a fixed distance $d=1.5\, \mathrm{m}$. We tested the denoising procedure considering values of $z$ in the range from $1$ and $q-n-1=14$. For all the $I$ Monte-Carlo runs, we generated all the possible combinations of $z$ TDOAs extracted from the last $n-q$ entries of the vector $\bs{\hat{\tau}}_i$. As before, we considered the LS, SRD-LS and GS algorithms for source localization. The results, averaged among all the noise realizations and all the combinations, are reported in Figure \ref{fig:test3}. In particular, Figure~\ref{fig:test3_TDOAerr} shows the residual error on TDOAs after denoising, while Figure~\ref{fig:test3_locRMSE} highlights the impact of denoising on localization. Note that we added to the graphs the points at $z=0$ (i.e., when only the $n$ TDOAs referred to the first microphone are available) and at $z=q-n=15$ (i.e., when all the TDOAs are used). It is worth noticing that the availability of just a few additional TDOAs leads to a relevant reduction of the TDOA standard deviation, with respect to $z=0$. This reflects also on localization, as all the algorithms monotonically improve their accuracy increasing the number of available measurements. Also in this case, GS exhibits the best accuracy after denoising, while being characterized by an unstable behavior using the original TDOAs. Indeed, with the original data GS improves its accuracy when $z<6$; for higher values of $z$, the error bias becomes relevant and the overall RMSE diverges. 

It is important to highlight the practical implications of these results. Let us consider a microphone array composed by $(n+1)$ sensors. The computational power requested for computing all the $q$ TDOAs is mainly due to the calculation of GCCs between the pairs of microphone signals. In case of limited computational capabilities, a typical solution would be that of computing only the nonredundant TDOA set. However, it turns to be more convenient to compute $z$ additional TDOAs in order to fully exploit the available computational power. This enables better localization, without modifying the array configuration.

\section{Conclusions}
\label{sec:conclusions}
In this manuscript we reformulated the problem of source localization in the TDOA space. This enabled us to show that source localization has a neat interpretation in terms of TDOA denoising. As a simple solution to the denoising problem is not available, we proved that it is possible to relax the problem to a linear one, whose solution is based on projecting TDOAs on a linear subspace of the TDOA space. Moreover, we also derived the problem solution for the case in which only a few TDOAs measurements are available.

The analysis performed in this manuscript does not limit to numerical simulations. Indeed, each choice behind the presented algorithm is fully justified by means of analytical proofs that further validate and justify the works in \citep{Cheung2008} and \citep{Schmidt1996}. As a matter of fact, in this manuscript we proved that denoising has a positive effect on source localization from a theoretical perspective. Moreover, we tested the denoising algorithm using different noise models, thus highlighting that the method is still valid even when noise hypotheses are not completely fulfilled. Finally, we also made use of different cost functions to gain an interesting insight on how denoising impacts on different localization algorithms.

The extension of the relaxed denoising algorithm to the case of missing TDOAs have also interesting implications in real-world scenarios. As a matter of fact, it enables to fully exploit hardware computational capabilities in order to increase localization performance. As an example, by fixing the available computational complexity, one can tune the localization system in order to measure a given amount of TDOAs and fully take advantage of them in a synergistic fashion.

According to the results of this work, it is important to note that it is possible to envision the development of algorithms working in the TDOA space to solve the complex ML source localization problem in a easier way. This will be the scope of possible future works.


\bibliographystyle{myspbasic}      
\bibliography{biblio}

\begin{thebibliography}{37}
\providecommand{\natexlab}[1]{#1}
\providecommand{\url}[1]{{#1}}
\providecommand{\urlprefix}{URL }
\expandafter\ifx\csname urlstyle\endcsname\relax
  \providecommand{\doi}[1]{DOI~\discretionary{}{}{}#1}\else
  \providecommand{\doi}{DOI~\discretionary{}{}{}\begingroup
  \urlstyle{rm}\Url}\fi
\providecommand{\eprint}[2][]{\url{#2}}

\bibitem[{Alameda-Pineda and Horaud(2014)}]{Pineda2014}
Alameda-Pineda X., Horaud R. (2014). A geometric approach to sound source
  localization from time-delay estimates. \emph{IEEE Transactions on Audio,
  Speech, and Language Processing} 22:1082--1095,
  \doi{10.1109/TASLP.2014.2317989}

\bibitem[{Amari and Nagaoka(2000)}]{Amari2000}
Amari S., Nagaoka H. (2000). \emph{Methods of Information Geometry}. American
  Mathematical Society

\bibitem[{Basu et~al(2006)Basu, Pollack, and Roy}]{Basu2006}
Basu S., Pollack R., Roy M. (2006). \emph{Algorithms in real algebraic
  geometry. Second edition.} Algorithms and Computation in Mathematics,
  Springer Verlag, Berlin, \doi{10.1007/3-540-33099-2}

\bibitem[{Beck et~al(2008)Beck, Stoica, and Li}]{Beck2008}
Beck A., Stoica P., Li J. (2008). Exact and approximate solutions of source
  localization problems. \emph{IEEE Transactions on Signal Processing}
  56:1770--1778, \doi{10.1109/TSP.2007.909342}

\bibitem[{Benesty and Huang(2004)}]{Benesty2004}
Benesty J., Huang Y. (eds)  (2004). \emph{Audio Signal Processing for
  Next-Generation Multimedia Communication Systems}. Springer

\bibitem[{Bestagini et~al(2013)Bestagini, Compagnoni, Antonacci, Sarti, and
  Tubaro}]{Bestagini2013}
Bestagini P., Compagnoni M., Antonacci F., Sarti A., Tubaro S. (2013).
  Tdoa-based acoustic source localization in the space--range reference frame.
  \emph{Multidimensional Systems and Signal Processing}
  \doi{10.1007/s11045-013-0233-8}

\bibitem[{Canclini et~al(2013)Canclini, Antonacci, Sarti, and
  Tubaro}]{Canclini2013}
Canclini A., Antonacci E., Sarti A., Tubaro S. (2013). Acoustic source
  localization with distributed asynchronous microphone networks. \emph{IEEE
  Transactions on Audio, Speech, and Language Processing} 21(2):439--443,
  \doi{10.1109/TASL.2012.2215601}

\bibitem[{Canclini et~al(2015)Canclini, Bestagini, Antonacci, Compagnoni,
  Sarti, and Tubaro}]{Canclini2015}
Canclini A., Bestagini P., Antonacci F., Compagnoni M., Sarti A., Tubaro S.
  (2015). A robust and low-complexity source localization algorithm for
  asynchronous distributed microphone networks. \emph{IEEE/ACM Transactions on
  Audio, Speech, and Language Processing} 23(10):1563--1575,
  \doi{10.1109/TASLP.2015.2439040}

\bibitem[{Chen et~al(2002)Chen, Hudson, and Yao}]{Chen2002}
Chen J., Hudson R., Yao K. (2002). Maximum-likelihood source localization and
  unknown sensor location estimation for wideband signals in the near-field.
  \emph{IEEE Transactions on Signal Processing} 50(8):1843 --1854,
  \doi{10.1109/TSP.2002.800420}

\bibitem[{Compagnoni and Notari(2014)}]{Compagnoni2013b}
Compagnoni M., Notari R. (2014). {TDOA–-based localization in two dimension:
  the bifurcation curve}. \emph{Fundamenta Informaticae} 135:199--210

\bibitem[{Compagnoni et~al(2012)Compagnoni, Bestagini, Antonacci, Sarti, and
  Tubaro}]{Compagnoni2012}
Compagnoni M., Bestagini P., Antonacci F., Sarti A., Tubaro S. (2012).
  Localization of acoustic sources through the fitting of propagation cones
  using multiple independent arrays. \emph{IEEE Transactions on Audio, Speech,
  and Language Processing} 20:1964--1975, \doi{10.1109/TASL.2012.2191958}

\bibitem[{Compagnoni et~al(2014)Compagnoni, Notari, Antonacci, and
  Sarti}]{Compagnoni2013a}
Compagnoni M., Notari R., Antonacci F., Sarti A. (2014). A comprehensive
  analysis of the geometry of tdoa maps in localization problems. \emph{Inverse
  Problems} 30(3):035,004,
  \urlprefix\url{http://stacks.iop.org/0266-5611/30/i=3/a=035004}

\bibitem[{D'Arca et~al(2014)D'Arca, Robertson, and Hopgood}]{D_Arca2014}
D'Arca E., Robertson N., Hopgood J. (2014). Look who's talking: Detecting the
  dominant speaker in a cluttered scenario. \emph{IEEE International Conference
  on Acoustics, Speech and Signal Processing, ICASSP '14}

\bibitem[{Gillette and Silverman(2008)}]{Gillette2008a}
Gillette M., Silverman H. (2008). A linear closed-form algorithm for source
  localization from time-differences of arrival. \emph{IEEE Signal Processing
  Letters (SPL)} 15:1--4

\bibitem[{Grafarend and Shan(2002)}]{Grafarend2002}
Grafarend E., Shan J. (2002). {GPS Solutions: Closed Forms, Critical and
  Special Configurations of P4P}. \emph{GPS Solutions} 5(3):29--41

\bibitem[{Hahn and Tretter(1973)}]{Hahn1973}
Hahn W., Tretter S. (1973). Optimum processing for delay-vector estimation in
  passive signal arrays. \emph{IEEE Transactions on Information Theory}
  19:608--614, \doi{10.1109/TIT.1973.1055077}

\bibitem[{Hu and Li(2002)}]{Hen2002}
Hu H.~H., Li D. (2002). Energy based collaborative source localization using
  acoustic micro-sensor array. \emph{2002 IEEE Workshop on Multimedia Signal
  Processing}, pp. 371--375

\bibitem[{Hu and Yang(2010)}]{Hu2010}
Hu J., Yang C. (2010). Estimation of sound source number and directions under a
  multisource reverberant environment. \emph{EURASIP Journal on Advances in
  Signal Processing} 2010:63

\bibitem[{Huang et~al(2001)Huang, Benesty, Elko, and Mersereati}]{Huang2001}
Huang Y., Benesty J., Elko G., Mersereati R. (2001). Real-time passive source
  localization: a practical linear-correction least-squares approach.
  \emph{IEEE Transactions on Speech and Audio Processing} 9:943--956,
  \doi{10.1109/89.966097}

\bibitem[{Ianniello(1982)}]{Ianniello1982}
Ianniello J. (1982). Time delay estimation via cross-correlation in the
  presence of large estimation errors. \emph{IEEE Transactions on Acoustics,
  Speech and Signal Processing} 30:998--1003

\bibitem[{Kehu et~al(2009)Kehu, Gang, and Luo}]{Kehu2009}
Kehu Y., Gang W., Luo L. Z.-Q. (2009). Efficient convex relaxation methods for
  robust target localization by a sensor network using time differences of
  arrivals. \emph{IEEE Transactions on Signal Processing} 57(7):2775--2784

\bibitem[{Knapp and Carter(1976)}]{Knapp1976}
Knapp C., Carter G. (1976). The generalized correlation method for estimation
  of time delay. \emph{IEEE Transactions on Acoustics, Speech and Signal
  Processing} 24:320--327

\bibitem[{Koch and Westphal(1995)}]{Koch95}
Koch V., Westphal R. (1995). New approach to a multistatic passive radar sensor
  for air/space defense. \emph{Aerospace and Electronic Systems Magazine, IEEE}
  10(11):24--32

\bibitem[{Lehmann and Casella(1998)}]{Casella1998}
Lehmann E.~L., Casella G. (1998). \emph{Theory of Point Estimation}, 2nd edn.
  Springer, New-York

\bibitem[{Schau and Robinson(1987)}]{Schau1987}
Schau H., Robinson A. (1987). Passive source localization employing
  intersecting spherical surfaces from time-of-arrival differences. \emph{IEEE
  Transactions on Acoustics, Speech and Signal Processing} 35:1223--1225,
  \doi{10.1109/TASSP.1987.1165266}

\bibitem[{Scheuing and Yang(2006)}]{Scheuing2006}
Scheuing J., Yang B. (2006). Disambiguation of {TDOA} estimates in multi-path
  multi-source environments {(DATEMM)}. \emph{IEEE International Conference on
  Acoustics, Speech and Signal Processing, ICASSP '06}, vol~4, pp. IV --IV,
  \doi{10.1109/ICASSP.2006.1661099}

\bibitem[{Scheuing and Yang(2008)}]{Scheuing2008}
Scheuing J., Yang B. (2008). Disambiguation of {TDOA} estimation for multiple
  sources in reverberant environments. \emph{IEEE Transactions on Audio,
  Speech, and Language Processing} 16:1479--1489,
  \doi{10.1109/TASL.2008.2004533}

\bibitem[{Schmidt(1972)}]{Schmidt1972}
Schmidt R. (1972). A new approach to geometry of range difference location.
  \emph{IEEE Transactions on Aerospace and Electronic Systems} AES-8:821--835,
  \doi{10.1109/TAES.1972.309614}

\bibitem[{Schmidt(1996)}]{Schmidt1996}
Schmidt R. (1996). Least squares range difference location. \emph{IEEE
  Transactions on Aerospace and Electronic Systems} AES-32:234--242,
  \doi{10.1109/7.481265}

\bibitem[{Smith and Abel(1987)}]{Smith1987}
Smith J., Abel J.~S. (1987). Closed-form least-squares source location
  estimation from range-difference measurements. \emph{IEEE Trans Acoust,
  Speech, Signal Processing} ASSP-35:1661--1669

\bibitem[{So et~al(2008)So, Chan, and Chan}]{Cheung2008}
So H.~C., Chan Y.~T., Chan F. K.~W. (2008). Closed-form formulae for
  time-difference-of-arrival estimation. \emph{IEEE Transactions on Signal
  Processing} 56:2614--2620

\bibitem[{Spencer(2007)}]{Spencer2007}
Spencer S. (2007). The two-dimensional source location problem for time
  differences of arrival at minimal element monitoring arrays. \emph{Journal of
  the Acoustical Society of America} 121:3579--3594

\bibitem[{Stoica and Nehorai(1988)}]{Stoica1988}
Stoica P., Nehorai A. (1988). {MUSIC}, maximum likelihood and cramer-rao bound.
  \emph{IEEE International Conference on Acoustics, Speech and Signal
  Processing, ICASSP '88}, \doi{10.1109/ICASSP.1988.197097}

\bibitem[{Teunissen(2000)}]{Teunissen2000}
Teunissen P. (2000). \emph{Adjustment Theory: An Introduction}. Series on
  mathematical geodesy and positioning, Delft University Press

\bibitem[{Trifa et~al(2007)Trifa, Koene, Moren, and Cheng}]{Trifa2007}
Trifa V., Koene A., Moren J., Cheng G. (2007). Real-time acoustic source
  localization in noisy environments for human-robot multimodal interaction.
  \emph{IEEE International Symposium on Robot and Human interactive
  Communication, RO-MAN '07}

\bibitem[{Valenzise et~al(2007)Valenzise, Gerosa, Tagliasacchi, Antonacci, and
  Sarti}]{Valenzise2007}
Valenzise G., Gerosa L., Tagliasacchi M., Antonacci F., Sarti A. (2007). Scream
  and gunshot detection and localization for audio-surveillance systems.
  \emph{IEEE Conference on Advanced Video and Signal Based Surveillance, AVSS
  '07}

\bibitem[{Yimin et~al(2008)Yimin, Amin, and Ahmad}]{Yimin2008}
Yimin Z., Amin M., Ahmad F. (2008). Localization of inanimate moving targets
  using dual-frequency synthetic aperture radar and time-frequency analysis.
  \emph{IEEE International Geoscience and Remote Sensing Symposium, 2008.
  IGARSS 2008}, vol~2, pp. II--33--II--36

\end{thebibliography}



\end{document}